\documentclass[twocolumn]{aastex631}
\usepackage{booktabs}
\usepackage{amsmath, mathtools,xparse}

\submitjournal{ApJ}
\shorttitle{Leaving No Branches Behind}
\shortauthors{Chuang et al.}

\begin{document}

\title{Leaving No Branches Behind: Predicting Baryonic Properties of Galaxies from Merger Trees}

\author[0000-0003-2069-9413]{Chen-Yu Chuang}
\affiliation{Institute of Astronomy, National Tsing Hua University, Hsinchu 30013, Taiwan}
\affiliation{Institute of Astronomy and Astrophysics, Academia Sinica,   Taipei 10617, Taiwan}

\author[0000-0002-8896-6496]{Christian Kragh Jespersen}
\affiliation{Department of Astrophysical Sciences, Princeton University, Princeton, NJ 08544, USA}

\author[0000-0001-7146-4687]{Yen-Ting Lin}
\affiliation{Institute of Astronomy and Astrophysics, Academia Sinica, Taipei 10617, Taiwan}
\affiliation{Institute of Physics, National Yang Ming Chiao Tung University, Hsinchu 30010, Taiwan}
\affiliation{Graduate Institute of Astrophysics, National Taiwan University, Taipei 10617, Taiwan}

\author{Shirley Ho}
\affiliation{Center for Computational Astrophysics, Flatiron Institute, New York, NY 10010, USA}
\affiliation{Department of Astrophysical Sciences, Princeton University, Princeton, NJ 08544, USA}
\affiliation{Department of Physics, Carnegie Mellon University, Pittsburgh, PA 15217, USA}

\author[0000-0002-3185-1540]{Shy Genel}
\affiliation{Center for Computational Astrophysics, Flatiron Institute, New York, NY 10010, USA}
\affiliation{Columbia Astrophysics Laboratory, Columbia University, 550 West 120th Street, 
New York, NY 10027, USA}

\begin{abstract}

Galaxies play a key role in our endeavor to understand how structure formation proceeds in the Universe.  For any precision study of cosmology or galaxy formation, there is a strong demand for huge sets of realistic mock galaxy catalogs, spanning cosmologically significant volumes. For such a daunting task, methods that can produce a direct mapping between dark matter halos from dark matter-only simulations and galaxies are strongly preferred, as producing mocks from full-fledged hydrodynamical simulations or semi-analytical models is too expensive. Here we present a Graph Neural Network-based model that is able to accurately predict key properties of galaxies such as stellar mass, $g-r$ color, star formation rate, gas mass, stellar metallicity and gas metallicity, purely from dark matter properties extracted from halos along the full assembly history of the galaxies.  Tests based on the {\it TNG300} simulation of the {\it IllustrisTNG} project show that our model can recover the baryonic properties of galaxies to high accuracy,  over a wide redshift range ($z = 0-5$), for all galaxies with stellar masses more massive than $10^9\,M_\odot$ and their progenitors, with strong improvements over the state-of-the-art methods. We further show that our method makes substantial strides towards providing an understanding of the implications of the {\it IllustrisTNG} galaxy formation model.

\end{abstract}


\section{Introduction} \label{sec:intro}
According to the standard model of cosmology, the matter content of the Universe is dominated by dark matter, which interacts with baryons mainly through gravity (\citealt{planck16}). Galaxies are believed to form in dark matter halos, which begin as low-mass entities and grow in mass by mergers and accretion of matter (e.g., \citealt{white78,springel05}). The hierarchical growth for any given dark matter halo can be visualized as a ``merger tree'' (see Figure \ref{fig:tree}).  It is believed that the halo growth history plays an important role in shaping the galaxies they host (\citealt{somerville15,naab17}), but it is non-trivial to recover the baryonic properties of galaxies solely from merger trees \citep{jespersen2022}. Much more sophisticated models, such as semi-analytical models  \citep[SAMs;][]{kauffmann1993, somerville2008, croton2016} or hydrodynamical simulations \citep[][]{pillepich2018, navarro2021, pakmor2022} are required, at the cost of several tens of million CPU (central processing unit) hours, which makes many otherwise valuable applications impossible.

For large-scale sky surveys aiming at understanding the constituents of the Universe, having realistic mock galaxy catalogs is essential, as they are {\it required} for estimating covariance matrices for constraining cosmological parameters, performing end-to-end validation of the analysis pipeline, as well as estimating the accuracy of photometric redshifts and masses of galaxies, just to name a few (e.g., \citealt{lsst22}). Cosmological surveys are particularly demanding, as tens of thousands of mocks over significant volumes are needed to reduce systematic uncertainties (e.g., \citealt{Villaescusa20,rossi21}). In order to efficiently generate realistic mock catalogs, many astronomers have resorted to machine learning (ML), and trained models with outputs from numerical simulations that can reproduce key properties of the galaxy populations, such as stellar mass ($M_\star$), color, and star formation rate (SFR). Previous works \citep[e.g.,][]{harshil2016, shankar2018, natali2022, christopher2022} have attempted to assign baryonic properties to dark matter halos through various ML algorithms, such as Multi-Layer Perceptrons (MLPs), Random Forests, Extremely Randomized Trees, or a combination of these. However, these works only utilize the dark matter properties at $z=0$, and an inefficient encoding of the merger history, leading to suboptimal performance. Recently, \citet{jespersen2022} have developed a Graph Neural Network (GNN) model, \verb|Mangrove|, to leverage the full information of merger trees. \citet{jespersen2022} recovered $M_\star$, cold gas mass and metallicity, and supermassive black hole mass with much higher precision than approaches not taking the merger history explicitly into account. However, the original version of \verb|Mangrove| only predicts the baryonic properties for a single galaxy per merger tree, and is based on only the Santa Cruz SAM \citep[][]{somerville2015}, instead of a cosmological hydrodynamical simulation.

In this paper, we put forth a model based on \verb|Mangrove| that uses all the available information of dark matter halos along the merger trees of a large sample of magnetohydrodynamically simulated galaxies, such that the relevant, {\it baryonic} properties of the model galaxies can be predicted to higher accuracy than ever before for all subhalos at $z\lesssim 5$, which has never been achieved before. 

This paper is structured as the following. In Section \ref{sec:data}, we describe how the data is obtained, the structure of the GNN model, and the loss function  utilized. In Section \ref{sec:res}, we present the overall performance of the model. Section \ref{sec:comp} compares the performance of our model with existing studies, while Section \ref{sec:exam} shows the prediction of different baryonic properties of galaxies. In Section \ref{sec:fin}, we discuss the caveats,  prospects of applications and potential extensions of our model.

Throughout this work, we adopt the same cosmological parameters as {\it TNG300} of the \textit{IllustrisTNG} project \citep[][hereafter TNG300]{Marinacci2018,Naiman2018,Nelson18,pillepich2018,springel18,Nelson2019}\footnote{$H_0=100h\,\rm{km\,s^{-1}\,Mpc^{-1}}$, with $h=0.6774$}.

\section{Data Preparation and Model}
\label{sec:data}
We start by describing the simulation data set used in this work (Section~\ref{sec:sim}), detailing our treatment of dark matter halo merger trees (Section~\ref{sec:datsel}).  We then present our GNN model (Section~\ref{sec:gnn}), describing the arrangement of our edge, node, and global attributes, the loss function as well as the metric used to gauge the model performance.

\subsection{Simulation and Merger Trees}
\label{sec:sim}
\begin{figure*}
  \centering
  \includegraphics[width=0.9\linewidth]{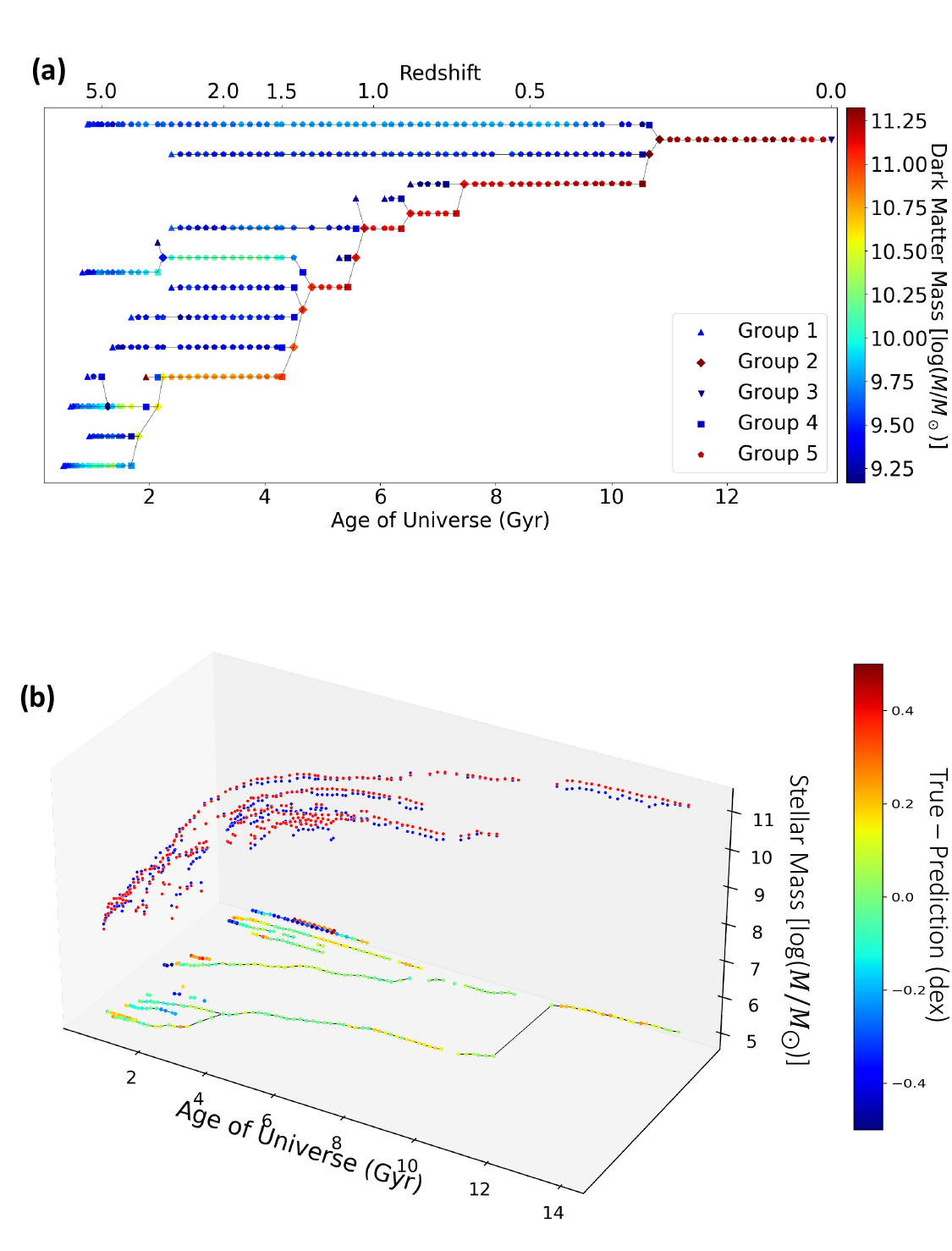}\caption{
  (a) An arbitrarily chosen merger tree depicting a subhalo with a final subhalo mass of $M_{\rm DM}\sim10^{11}\ \rm{M_\odot}$ (with $M_\star\sim10^{10}M_\odot$). The $x$-axis represents the age of the universe at each snapshot, while the color indicates the dark matter halo mass. The nodes are categorized into 5 groups (to be described in Section \ref{sec:rearr} and Figure \ref{fig:workflow}b), which correspond to different shapes in the figure. (b) A comparison of true (red data points) and predicted (blue data points) logarithmic stellar masses (shown on the $z$-axis) in a sample merger tree with a final stellar mass of $\sim10^{11}\ \rm{M_\odot}$. The color in the merger tree on the $x-y$ plane indicates the prediction error, measured in dex. Our work mainly focuses on the subhalos with $M_{DM}\geq 10^{11}h^{-1}\,M_\odot$, and thus, the other subhalos are omitted.}
  \label{fig:tree}
\end{figure*}

To train our model, we extract the dark matter and baryonic features from the TNG300, which has a box of $303\,\rm{Mpc}$ on a side and contains $(2500)^3$ dark matter particles. The gravitationally bounded subhalos (substructures of halos; one can regard that galaxies all live in subhalos) are identified by the \texttt{SUBFIND} algorithm \citep{springel2001}. All the properties in this work are derived from individual subhalos and their respective parent halos. We extract the baryonic properties and dark matter properties for each subhalo from the full magnetohydrodynamical (MHD) simulation and the corresponding dark matter-only (DMO) simulation separately. The subhalo-level merger trees are constructed by the \texttt{Sublink} algorithm \citep{gomez2015}. Figure \ref{fig:tree} depicts an arbitrarily chosen merger tree with a final dark matter subhalo mass\footnote{Total mass of all member particle/cells which are bound to a subhalo.} of $M_{\rm DM} \sim 10^{11}\ h^{-1}\rm{M_\odot}$

\subsection{Data Selection and Augmentation}
\label{sec:datsel}
In this work, we aim to reconstruct the stellar mass, SFR, $g - r$ color, gas mass ($M_{\rm gas}$), gas metallicity ($Z_{\rm gas}$) and the stellar metallicity ($Z_\star$) along the whole merger tree. These are key quantities that can be inferred from the observables of galaxies and are therefore important for creating mock catalogs. We limit ourselves to using merger trees that have a final dark matter subhalo mass of $M_{DM}\geq 10^{11}h^{-1}\,M_\odot$ and with a total number of subhalos throughout the tree of between $10-2\times10^4$, but we validate, test, and show the results (Section \ref{sec:res}) only for subhalos at a redshift $z\leq2$ (when the age of the universe is $\geq3.285\,\rm{Gyr}$) and with a dark matter mass $\geq 10^{11}h^{-1}\,M_\odot$.\footnote{We choose this mass limit as it corresponds to the population of galaxies with $M_\star\gtrsim10^9h^{-1}\,M_\odot$, a threshold limit relevant to most current observations at $z\lesssim 2$.} 

For the dark matter features, we use all subhalo and halo features provided by the TNG300 simulation except for the IDs as well as 3D positions, 3D velocities, and 3D spin. Although the $xyz$-components spin is defined with respect to the halo and thus formally invariant to changes in the coordinate system, we decided to exclude it to be sure that no spatial information leakage between the training and test set would occur. However, because the total spin $J = \sqrt{J_x^2+J_y^2+J_z^2}$ should be spatially uncorrelated, we decide to include this. Since merger trees consist of subhalos (which in a graph can be abstracted  as nodes) and the links (which in a graph can be abstracted  as edges) between subhalos at different snapshots, we also utilize features for the edges (i.e., mass ratio of a progenitor and a descendent linked by an edge). Lastly, we include global features of the merger trees (e.g., the toal number of progenitors in merger trees). Please refer to Appendix \ref{sec:fea} for the full list of the dark matter features used. 

We extract the baryonic properties for each subhalo from the full  MHD simulation and map the properties to the subhalos in the merger trees of the DMO simulation using the \texttt{LHaloTree} algorithm \citep{nelson2015}. Bijective matching of subhalos is achieved by comparing unique dark matter particle IDs, matching the subhalos with the highest fractions of common particles. At $z=0$, the matching fraction of the subhalos with $M_{DM}\geq 10^{11}h^{-1}\,M_\odot$ is $100\%$, while the matching fraction of subhalos in the merger trees used for training, validation, and testing our model throughout cosmic history is $73.14\%$. During the classification process described in Section \ref{sec:loss}, we assume that non-matched subhalos in the DMO simulation do not contain baryonic components and they are therefore omitted in the regression process.

\subsubsection{Data Split}
\label{sec:dataspl}
To prevent information leakage between the training, validation, and test data sets, we divide the TNG300 box into 8 equal-sized sub-boxes. Out of these sub-boxes, 6 are allocated to the training set, one to the validation set, and one to the test set. We randomly sample the 7231 merger trees in the training set, 1057 in the validation set, and 1031 in the test set. We train the GNN model using different combinations of hyperparameters and verify its performance on the validation set. We select the best-performing model and present the results obtained from that model on the test set in the following sections. The test set is never used for model tuning, and is never used before the final results are prepared. For details regarding the tested hyperparameters, please refer to Appendix \ref{sec:hyp}.

\subsection{Graph Neural Networks}
\label{sec:gnn}
Our main goal is to map the properties of galaxies onto merger trees, which can easily be represented as \textit{directed graphs}. Thus, it is natural to employ a GNN to extract information from merger trees. Merger trees exhibit intricate structures, yet they can be handled through three fundamental graph components: nodes ($V$), edges ($E$), and global attributes ($U$). Nodes are entities with attributes, such as dark matter subhalos with a radius and mass. Edges describe the relation among different subhalos at different snapshots including a merger or inheritance. Edges can have attributes as well, such as the mass ratio of a progenitor and a descendent. The ratio could indicate which progenitor has a larger impact on the descendent. The relation between the baryonic and dark matter properties of a subhalo can be sensitive to the global condition of a merger tree, such as the total number of progenitors, which is then tracked as a global attribute. 

Typically, a GNN layer consists of a sequence of message passing, aggregation, and node/edge/global update processes. In our work, a merger tree is provided as an input to the GNN in the form of a graph. Each node aggregates information from its neighboring nodes and updates its own state accordingly. A GNN layer can be divided into three distinct subcomponents: the edge model ($\phi^e$), the node model ($\phi^v$), and the global model ($\phi^u$). In our implementation, all of these submodels consist of two layers of MLPs and include a ReLU activation function between each MLP layer. The edge model takes the features from each pair of connected nodes ($\mathbf{v_i},\ \mathbf{v_j}\in\mathbb{R}^{2L_v}$), where $L_v$ represents the number of features in each node, along with the edge attributes, to generate a message vector. The message vectors associated with each target node are element-wise aggregated and passed to the node model. The node model then updates the features of the target nodes based on the aggregated messages, their original features, and the global properties of the graph. Finally, the global model aggregates all the updated node features and message vectors to update the global properties of the graph.

Previous works have developed different kinds of GNN layers for different scientific purposes. Among them, \citet{battaglia2018} developed the \texttt{MetaLayer}, a flexible type of GNN layer that allows for the incorporation of any relational inductive bias. We customize the \texttt{MetaLayer} to study the mapping between dark matter merger trees and galaxies. We introduce a strong inductive bias that assumes merger events are significant in the growth of dark matter subhalos. We implement the inductive bias by rearranging the edge connection as described in Section \ref{sec:rearr}. In our full model, we utilize four sequential \texttt{MetaLayers}. The model is trained by minimizing the loss described in Section \ref{sec:loss} while we gauge the performance of the model with several different metrics described in Section \ref{sec:metrics}. A schematic view of the information flow within the GNN during training is provided in Figure~\ref{fig:workflow}a. 

\subsubsection{Rearrangement of Edge Connections}
\label{sec:rearr}
\begin{figure*}
  \centering
  \includegraphics[width=0.75\linewidth]{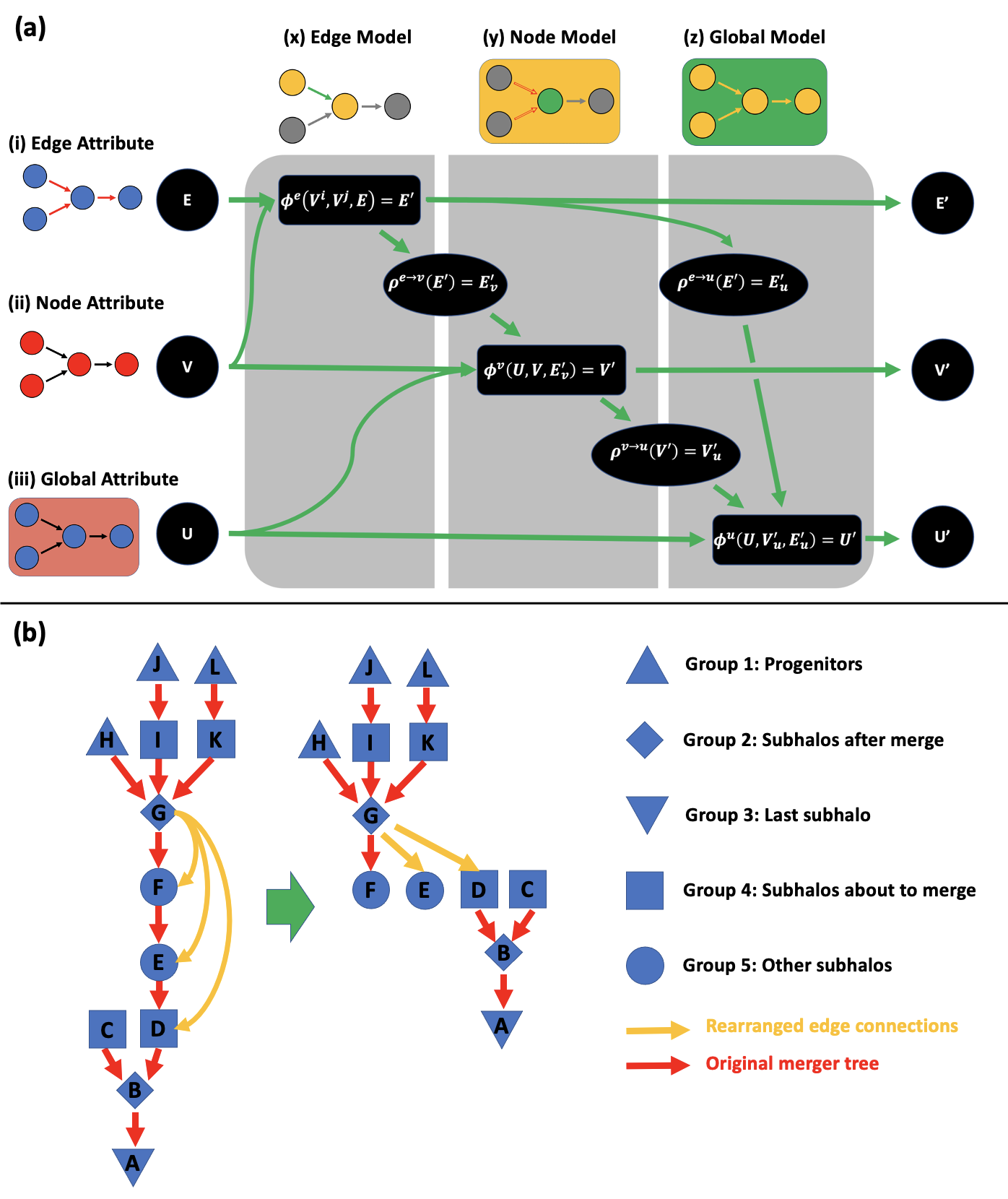}\caption{
  (a) A schematic view of the GNN workflow in a single \texttt{MetaLayer}. Merger trees exhibit intricate structures, yet they can be handled through three fundamental graph components: nodes ($V$), edges ($E$), and global attributes ($U$). The red parts in rows (i), (ii), and (iii) indicate the node, edge, and global features. When a graph is processed through a \texttt{MetaLayer}, three learnable update functions ($\phi^e$, $\phi^v$, and $\phi^u$) and three aggregation functions ($\rho^{e\rightarrow v}$, $\rho^{e\rightarrow u}$, and $\rho^{v\rightarrow u}$) are applied to the input graph. The computation proceeds from the edge to the node and, finally, to the global level. Columns (x), (y), and (z) indicate the graph elements that are involved in each of these computations, respectively. The green color indicates which sub-model is being updated in each column, and the yellow color represents the additional elements that are involved in the update. The green arrows connect the three fundamental components (node, edge, and global attributes) and the sequence of functions applied to them. The process shows the following steps: (1) $\phi^e$ is applied edge-wise, which takes the attributes from the upstream nodes ($V^i$), the downstream nodes ($V^j$), and the edge itself ($E$), and returns an updated version of the edge attribute ($E'$). (2) $\rho^{e\rightarrow v}$ is applied for each node, which takes all $E'$s that project to the node and outputs $E'_v$. (3) $\phi^v$ is applied node-wise by taking $U$, $V$ and $E'_v$ as the input and returning the updated node attributes ($V'$). (4) $\rho^{e\rightarrow u}$ and $\rho^{v\rightarrow u}$ are applied to all the edges and nodes in the graph, to yield globally summarized edge $E'_u$ and node ($V'_u$) properties, respectively. (5) Finally, $\phi^u$ is applied to each full graph, taking $E'_u$, $V'_u$, and $U$ as the input to update the global attribute ($U'$). Now, we have the updated edge, node and global properties for the graph.
  \\
  (b) A simple merger tree, as described in Section \ref{sec:sim}, to illustrate the connections between different nodes. The red arrows denote the original merger tree, while the yellow arrows depict the rearranged edge connections described in Section \ref{sec:rearr}. The rearranged edges allow the latest merger events to have added impact on all subhalos following a merger. Yet, by the rearrangement, we introduce a strong inductive bias that assumes merger events are significant in the growth of dark matter subhalos due to the removal of the smooth accretion mode in subhalos' growth (e.g. removing the connection between nodes F, E, and D).
  }
  \label{fig:workflow}
\end{figure*}

Given that the merger trees span 100 simulation snapshots, a GNN with 100 layers would  be required to transmit information from the beginning to the end of the merger trees. However, such an approach would result in over-smoothing, which negatively affects the predictive ability of individual nodes and the GNN as a whole \citep{oono2019}. To facilitate information transfer and impose an inductive bias of merger events having higher importance, we modify the connection scheme for each subhalo. First, we classify the subhalos into five categories (shown by different shapes of nodes in Figure \ref{fig:tree} and Figure \ref{fig:workflow}b):
\begin{itemize}
    \item (1) Progenitors (i.e., subhalos that are the first to be identified on a given branch).
    \item (2) Subhalos directly after merger events.
    \item (3) $z=0$ (i.e., the last) subhalo in the merger tree.
    \item (4) Subhalos that will merge in the next snapshot (i.e., pre-merger subhalos).
    \item (5) All other subhalos (e.g., isolated ones).
\end{itemize}
Next, we assign edges with the following rules: 
\begin{itemize}
    \item (a) Subhalos that are either pre-merger, last subhalo, or isolated (i.e., other subhalos) \{(3),(4),(5)\} are connected by the most recent post-merger or progenitor subhalo \{(1),(2)\}. 
    \item (b) Pre-merger subhalos \{(4)\} are connected to post-merger subhalos \{(2)\}.
    \item (c) Subhalos in \{(5)\} do not connect to any other subhalos.
\end{itemize}
Figure \ref{fig:workflow}b provides a schematic view of the rearrangement process. This reduces the number of required layers while preserving the inductive bias related to the importance of merger events. Nonetheless, there are some subhalos that conform to more than one of the five categories. For the subhalos that belong to both categories (1) and (4), (4) takes priority over (1) and will be connected to the post-merger subhalos. For the subhalos that conform to categories (2) and (3), we adopt (2) over (3) and the subhalos will be connected by those in (4). For the subhalos belonging simultaneously to  categories (2) and (4), both the rules applied to (2) and (4) take effect individually. The other conformations either do not exist by definition of the categories or are excluded by the tree selection procedure. 

\subsubsection{Loss function}
\label{sec:loss}
In this section, we define the loss function used to train our model. It comprises three distinct components: a {\it classification} loss to determine the presence of corresponding baryonic features in dark matter subhalos, a {\it regression} loss to predict the baryonic features, and {\it L1 and L2 norms} (defined in Equation~\ref{eq:ln}) to prevent overfitting.

Throughout different snapshots in the TNG300 dataset, an average of $\sim70\%$ of the dark matter subhalos do not contain stars, $\sim80\%$ lack star-formation activity, and $\sim40\%$ do not contain gas. Thus, our model is trained to simultaneously distinguish whether a subhalo contains stars, gas, and/or star-formation activity and to regress the amount in the case that a subhalo does contain a non-zero amount of the relevant target. We define subhalos containing star and gas particles or  are star-forming as positive cases, while those lacking  stars/gas or star formation are considered negative. The total loss therefore has a classification component, quantified using a cross-entropy loss, which is commonly used to measure the discrepancy between two discrete probability distributions 
\citep[][]{good1952},
\begin{equation}
\label{eq:cross}
\begin{aligned}
H(P,Q)_{\rm f,i}&=-\sum_{n=0}^{1} Q_n\log(P_n) \\
    &=-\left( q\log(p)+(1-q)\log(1-p) \right)
\end{aligned}
\end{equation}
where $H_{f, i}$ represents the cross-entropy loss for baryonic target $f$ of subhalo $i$, while $P_n$ and $Q_n$ denote the predicted and true probability, respectively, for a subhalo to be classified as class $n$ (i.e., be classified as either having a zero or non-zero amount of stars, gas, or star formation). Our model only consists of two classes (positive or negative), resulting in a summation with only two terms. In Equation~\ref{eq:cross}, $p$ and $q$ respectively represent the predicted and true probabilities for a subhalo to be positive 

For subhalos classified as containing star particles, we perform regression on the variables $M_\star$, color, and $Z_\star$. For those classified as containing gas particles, we regress the $M_{\rm gas}$ and $Z_{\rm gas}$. If a subhalo is classified as star-forming, we also regress the SFR. The regression loss is quantified with a Gaussian negative log-likelihood 
\begin{equation}
\label{eq:gauss}
G(y_{f,i}, \widehat{y_{f,i}}, \widehat{\sigma_{f,i}}) = \frac{1}{2}\left(2\log(\widehat{\sigma_{f,i}})+\frac{(y_{f,i}-\widehat{y_{f,i}})^2}{\widehat{\sigma_{f,i}}^2}\right),
\end{equation}
where $G$ represents the Gaussian negative log-likelihood loss, $y_{f,i}$ and $\widehat{y_{f,i}}$ denote the true and predicted values of baryonic feature $f$ for subhalo $i$, respectively. Additionally, $\widehat{\sigma_{f,i}}^2$ is the predicted variance of the baryonic feature $f$ for subhalo $i$. 

Lastly, we include the L1 and L2 norms of all parameters in the GNN in order to regularize the model and prevent overfitting, 
\begin{equation}
\label{eq:ln}
\begin{cases}
    \lVert N \rVert_{L_{1}}=\sum_{i\in {\rm GNN}}|P_i|\\
    \lVert N \rVert_{L_{2}}=\sum_{i\in {\rm GNN}}P_i^2
\end{cases}
\end{equation}
where $\lVert N \rVert_{L_{1}}$ and $\lVert N \rVert_{L_{2}}$ are the L1 and L2 norms, $P_i$ is the $i$th parameter in the GNN. 

Combining the losses above, the total loss we use to optimize the GNN is
\begin{equation}
\label{eq:tot}
L=\sum_{f,i}\left(H_{f,i} \cdot G_{f,i}+hH_{f,i}\right)+l_1\lVert N \rVert_{L_{1}}+l_2\lVert N \rVert_{L_{2}}
\end{equation}
where $h$, $l_1$, and $l_2$ are the weights assigned to the cross-entropy loss, L1 norm and L2 norm, respectively. The loss weights $h$, $l_1$, and $l_2$'s are found during the hyper-parameter search described in Appendix 
\ref{sec:hyp}.

\subsubsection{Metrics}
\label{sec:metrics}
To quantify the performance of our model, we introduce five metrics to gauge its prediction. For classification, we use the F1 score, a combined measure of precision and recall. Precision measures the proportion of correctly identified positive subhalos among all subhalos classified as positive (that is, $\text{Precision}=\frac{TP}{TP+FP}$, where $TP$ is the number of true positives and $FP$ is the number of false positives), while recall quantifies the model's ability to correctly predict positive subhalos out of all true positive subhalos (i.e., $\text{Recall}=\frac{TP}{TP+FN}$, where FN is the number of false negative results). The F1 score is then defined as the \textbf{harmonic mean} of the precision and recall:
\begin{equation} \label{eq:f1}
F1 = \frac{2}{\frac{1}{\text{Precision}}+\frac{1}{\text{Recall}}}
=\frac{TP}{TP+\frac{1}{2}(FP+FN)}
\end{equation}
A higher F1 score indicates a better classifier. The F1 is preferred over precision and recall since it is still a valuable metric for heavily imbalanced datasets.

We define four metrics for the regression following \citet{jespersen2022}. The first is the \textbf{scatter} of the prediction residuals,
\begin{equation} \label{eq:scat}
\sigma(y)=\sqrt{\frac{1}{N}\sum^N_i(\Delta y_i-\overline{\Delta y})^2}
\end{equation}
where $N$ is the number of galaxies in the test set, and $\Delta y_i=
\log(y_{\rm bar})-\log(\widehat{y_{\rm bar}})$ is the residual of a single prediction for a specific baryonic property in dex.\footnote{For color, $\Delta y_i \equiv
y_{\rm bar}-\widehat{y_{\rm bar}}$ as magnitudes  are already in log scale.} $\overline{\Delta y}$ is the average of the residual.
The second metric is the \textbf{bias}, defined as
\begin{equation} \label{eq:bias}
b(y)=\sum^N_i \frac{\Delta y_i}{N}.
\end{equation}
Since scatter and bias can be strongly affected by a few outliers, and is also directly included in the Gaussian regression loss, we include two additional metrics. 
Our third metric
is the Pearson correlation coefficient ($\rho$), which represents the linear correlation between the truth and the prediction of the model:
\begin{equation} \label{eq:pearson}
\rho=\frac{\text{cov}(y,\hat{y})}{\sigma_y\sigma_{\hat{y}}}=\frac{\sum_{i}^{N}(y_i-\bar{y})(\widehat{y_i}-\overline{\widehat{y}})}{\sqrt{\sum_i^N(y_i-\bar{y})^2\sum_i^N(\widehat{y_i}-\overline{\widehat{y}})^2}},
\end{equation}
where $y_i=\log(y_{\rm bar})$, $\widehat{y_i}=\log(\widehat{y_{\rm bar}})$, $\bar{y}$ is the mean of $y_i$, and $\bar{\hat{y}}$ is the mean of $\widehat{y_i}$. The last metric is the coefficient of determination ($R^2$), which represents the proportion of the variance in the predicted population $\hat{y}$ that can be explained by the true population $y$:
\begin{equation} \label{eq:r2}
R^2=1-\frac{\sum_i^N (\Delta y_i)^2}{\sum_i^N(y_i-\bar{y})^2}.
\end{equation}
A set of predictions all equal to the truth would result in $\rho=R^2=1$. 

\section{Results}
\label{sec:res}
We first compare the performance of our model with a couple of popular methods used in studying galaxy--halo connection (Section~\ref{sec:comp}), then describe in more details of the characteristics of our model prediction,
paying special attention to the stellar mass growth history (Section~\ref{sec:exam}).

\subsection{Comparison with Existing Methods}
\label{sec:comp}

We first compare our results with two other frameworks: an MLP that utilizes all information from the subhalo but lacks explicit merger information, and Abundance Matching \citep[AM, e.g.,][]{kravtsov04, Chuang2023}, which is widely used for connecting subhalo masses with stellar masses. AM can be performed based on subhalo mass, or with other quantities such as the peak maximum circular velocity a subhalo ever attained ($V_{\rm peak}$). Here, we compare two AM schemes, $V_{\rm max,90\%}$ and $\psi_5$, as presented in \citet{Chuang2023}, which are shown to be better tracers of stellar mass and mass-dependent two-point correlation function (2PCF) than $V_{\rm peak}$ and subhalo mass (please refer to the footnotes of Table \ref{tab:comp} for the definitions of the two AM schemes). Furthermore, we compare our results with the \textbf{chaotic uncertainty limit} introduced by \citet{genel2019}, which represents a theoretical absolute lower limit on the optimal performance achievable by a perfect predictor (an \textit{ideal machine}). The chaotic limit is found as the scatter between runs of simulations that only differ by an infinitesimal perturbation at $z=5$. The results of the three frameworks (MLP plus two AM schemes), along with the chaotic uncertainty limits for the predicted properties at $z = 0$, are presented in Table \ref{tab:comp}. The GNN, utilizing the merger trees, outperforms the MLP and AM in all predicted features, except for a negligible difference in the bias, which is always small. However, the improvement from MLP to GNN is only substantial (exceeding 10\%) for the scatters of $M_\star$, $M_{\rm gas}$ and $Z_\star$, while the improvements in other properties are relatively small. 

\begin{deluxetable*}{ll|llllll}
\tablecaption{
Comparison of different methods with four of metrics. The best performance is shown in bold for each metric. We also include the chaotic uncertainty limit, $\sigma_0$, which was introduced in Section \ref{sec:comp}. $\sigma_0$ is the best possible regression performance. The \texttt{Mangrove}-based GNN always outperforms the MLP, meaning that including the merger history always improves predictions, although not always significantly. All the results are unbiased. The improvements are similar at both $z=0$ and over the period of cosmic noon to the present. The small biases along with a more complete set of metrics can be found in Table \ref{tab:comp1} in  Appendix~\ref{app:fulmet}. Although our model achieves an accurate mapping between the dark matter halo merger history and some relevant baryonic properties of galaxies such as $M_\star$, $Z_\star$, and $M_{\rm gas}$, the predictions of SFR, color, and $Z_{\rm gas}$ still show a relatively large scatter compared to the true values. This can be attributed to the stochastic nature of the SFR, color, and $Z_{\rm gas}$ history. To characterize the influence of the stochastic components on different baryonic features, a comparison of the GNN prediction on the true and smoothed baryonic feature history is presented in Table \ref{tab:compsmth}. \label{tab:comp}}
\tablehead{
Target & Limit [$\sigma_0$] & Method & $\sigma$ & Improvement\tablenotemark{a} & $\rho$ & $R^2$ & F1\\
\hline
\multicolumn{8}{c}{redshift 0}}
\startdata
$M_*$& 0.103& GNN& \textbf{0.141}& \textbf{45\%} & \textbf{0.979}& \textbf{0.959}& \textbf{1.0}\\
& & AM$_{\psi_5}$ \tablenotemark{b}& 0.184 & 3\%&0.959& 0.920\\
& & AM$_{V_{\rm 90\%}}$ \tablenotemark{c}&0.222& -34\% & 0.963 & 0.927\\
& & MLP& 0.187& & 0.963& 0.928& \textbf{1.0}\\
\hline
SFR& 0.354& GNN& \textbf{0.388}& \textbf{13\%}& \textbf{0.768}& \textbf{0.59}& \textbf{0.968}\\
& &MLP& 0.4& & 0.75& 0.562& 0.964\\
\hline
color ($g-r$)& 0.118& GNN& \textbf{0.129}& \textbf{4\%}& \textbf{0.693}& \textbf{0.48}& \textbf{1.0}\\
& & MLP& 0.133& & 0.667& 0.444&\textbf{1.0}\\
\hline
$M_{\rm gas}$&0.132& GNN& \textbf{0.187}& \textbf{22\%}& \textbf{0.88}& \textbf{0.775}& \textbf{0.994}\\
& & MLP& 0.216& & 0.844& 0.713&\textbf{0.994}\\
\hline
$Z_{\rm gas}$&0.154& GNN& \textbf{0.179}& \textbf{5\%}& \textbf{0.773}& \textbf{0.597}& \textbf{0.991}\\
& & MLP& 0.186& & 0.749& 0.562&0.99\\
\hline
$Z_*$& 0.094& GNN& \textbf{0.111}& \textbf{13\%}& \textbf{0.918}& \textbf{0.843}& \textbf{1.0}\\
& & MLP& 0.123& & 0.9& 0.81& \textbf{1.0}\\
\hline
\hline
\multicolumn{8}{c}{redshift $0-2$}\\
\hline
$M_*$& & GNN& \textbf{0.145}&  & \textbf{0.979}& \textbf{0.958}& \textbf{0.999}\\
& & MLP& 0.184&  & 0.966& 0.934&0.998\\
\hline
SFR& & GNN& \textbf{0.331}& & \textbf{0.87}& \textbf{0.757}& \textbf{0.985}\\
& & MLP& 0.348&  & 0.859& 0.738& 0.983\\
\hline
color ($g-r$)& & GNN& \textbf{0.108}&  & \textbf{0.779}& \textbf{0.607}& \textbf{0.999}\\
& & MLP& 0.112& & 0.76& 0.577&0.998\\
\hline
$M_{\rm gas}$& &GNN& \textbf{0.127}&  & \textbf{0.942}& \textbf{0.888}& \textbf{0.998}\\
& & MLP& 0.143&  & 0.928& 0.862& 0.997\\
\hline
$Z_{\rm gas}$& & GNN& \textbf{0.158}&  & \textbf{0.861}& \textbf{0.774}& \textbf{0.996}\\
& & MLP& 0.17&  & 0.846& 0.715& \textbf{0.996}\\
\hline
$Z_*$& & GNN& \textbf{0.125}&  & \textbf{0.924}& \textbf{0.853}& \textbf{0.999}\\
& & MLP& 0.144&  & 0.905& 0.818& 0.998\\
\hline
\enddata
\tablenotetext{a}{Defined as $(\sigma_{\rm MLP}-\sigma_{\rm X})/\sigma_0$, where $X$ is the method (GNN or AM).}
\tablenotetext{b}{Abundance matching using $\psi_5\equiv\frac{V_{\rm max,90\%}}{V_{\rm max,90\%@13.2}}+\frac{|\dot{M}_{\rm DM}|_{\rm 60\%}}{|\dot{M}_{\rm DM}|_{\rm 60\%@13.2}}$, where the parameters with a subscript $\rm @$ are the normalization factor at a fitted pivot $M_{\rm DM,peak}$, $V_{\rm 90\%}$ is the 90th percentile for the maximum circular velocity ($V_{\rm max}$) throughout the lifetimes of subhalos, and $|\dot{M}_{\rm DM}|_{\rm 60\%}$ is the absolute subhalo dark matter mass variation rate at 60th percentile. }
\tablenotetext{c}{Abundance matching using $V_{\rm 90\%}$. }
\end{deluxetable*}

\begin{figure*}
  \centering
  \includegraphics[width=0.87\linewidth]{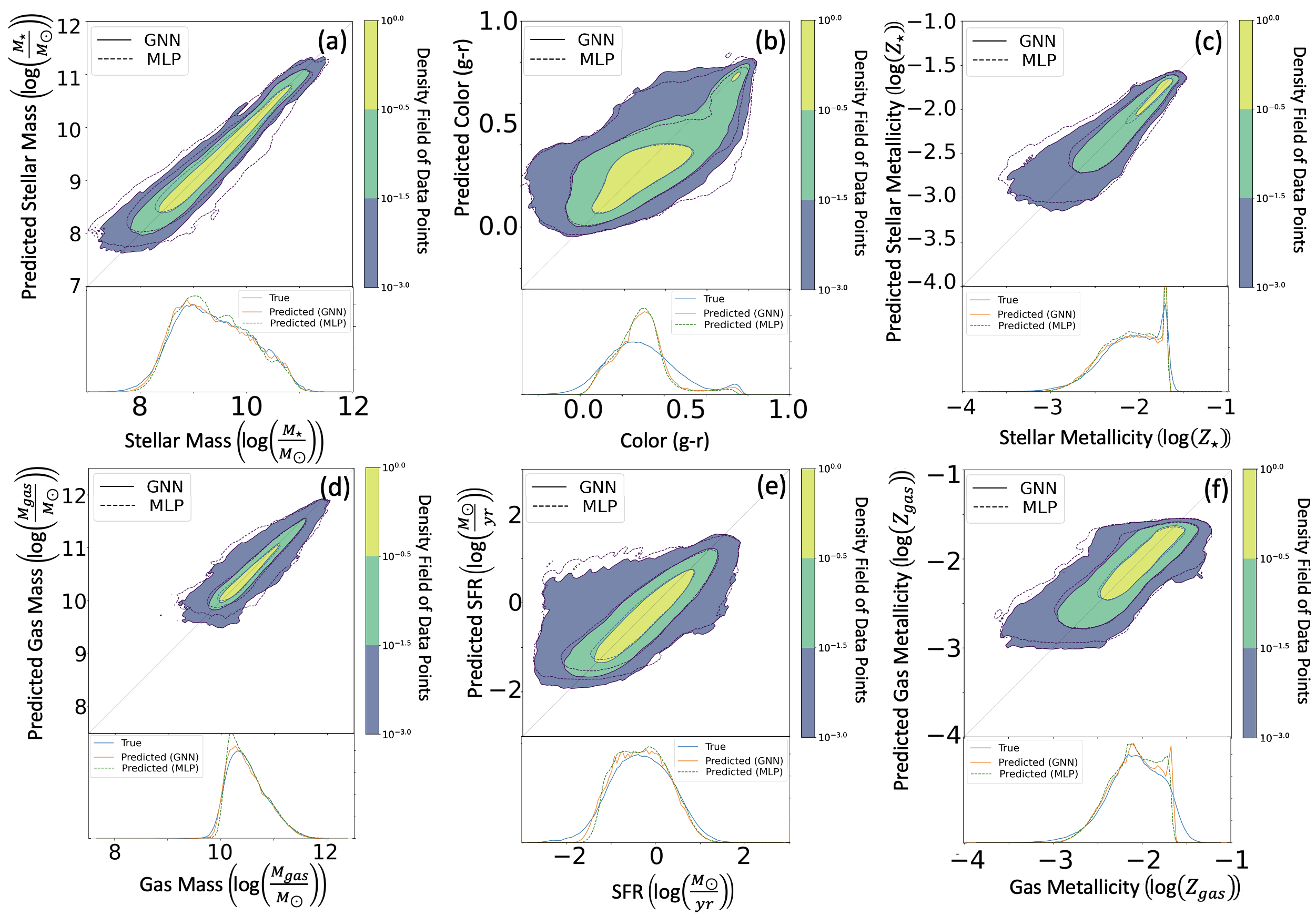}
  \caption{
  The upper part of each panel shows the comparison of predicted and true values of (a) stellar mass, (b) color ($g-r$), (c) stellar metallicity, (d) gas mass, (e) SFR, and (f) gas metallicity. The color scale indicates the density of the GNN-predicted data points, normalized such that the maximum value is one. The dashed and solid contours represent the MLP-predicted and GNN-predicted baryonic features, respectively. The improvement from an MLP to our GNN is always visible. The lower part of each panel displays the number of subhalos in the predicted and true baryonic feature bins, also normalized such that the maximum value is one for the true distribution. The solid blue lines indicate the true distribution, the solid orange lines represent the GNN-predicted distribution, and the dashed green lines correspond to the MLP-predicted distribution. The distributions mostly follow each other, but both ML methods encounter difficulties when a bimodality exists in the distribution.
  }
  \label{fig:allfeat}
\end{figure*}

\begin{figure*}
  \centering
  \includegraphics[width=0.87\linewidth]{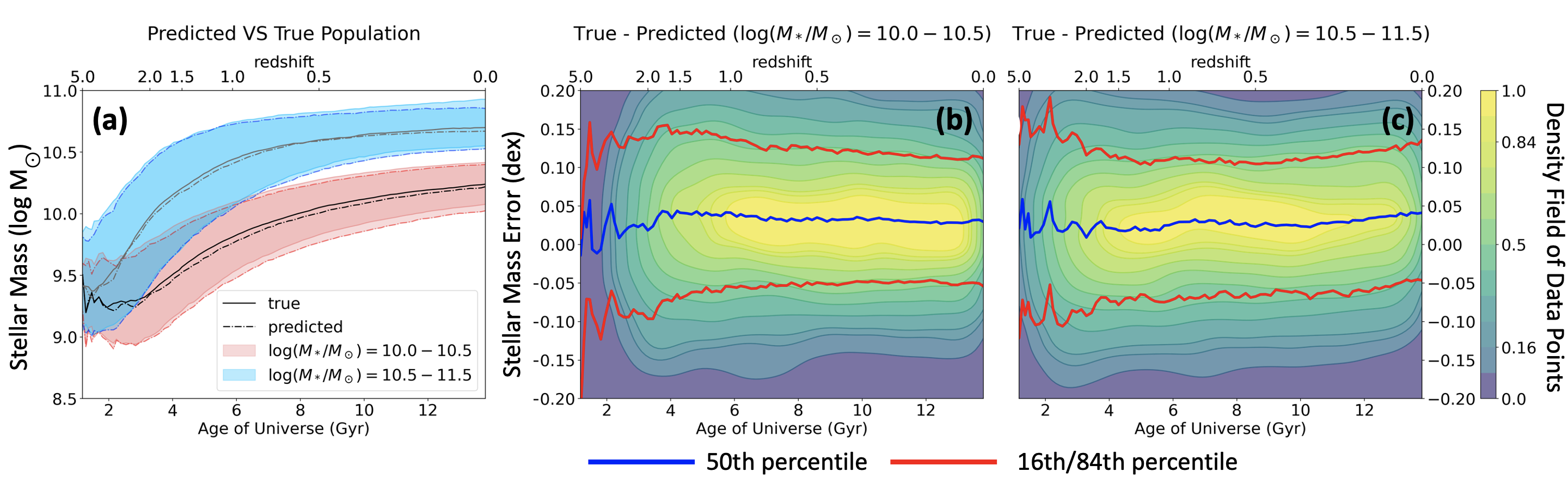}
  \caption{Panel (a): The stellar mass growth history in the primary branches of subhalos, for two stellar mass bins chosen at $z=0$ as indicated in the inset. The $y$-axis shows the stellar mass in the log-scale. The solid lines represent the true values, while the dash-dotted lines correspond to the GNN-predicted 16th, 50th, and 80th percentiles for the two mass bins. Panels (b) and (c): A detailed view of the difference between the true and the GNN-predicted stellar mass growth history. The color scale indicates the density of data points in different mass bins (indicated at the top of each panel), normalized such that the maximum value is one. The red lines indicate the 16th and 84th percentiles of the error at each redshift, while the blue ones represent the median error.}
  \label{fig:masshist}
\end{figure*}

\subsection{Analysis of the Predictions}
\label{sec:exam}
Figure \ref{fig:allfeat} shows the true and predicted values of all subhalos with $M_{DM}\geq10^{11}h^{-1}\, \rm{M_\odot}$ and at $z\leq2$ for both GNN and MLP models. For both models, the scatter in the predictions of $M_\star$ and $Z_{\rm gas}$ does not vary significantly with the absolute values of these properties. However, the scatter in the predictions of SFR, $Z_{\rm *}$ and $M_{\rm gas}$ tends to be larger at lower values. The scatter of color tends to be larger for galaxies in between the red sequence and blue cloud (i.e., in the green valley). Comparing the predictions of GNN and MLP, the GNN shows a greater reduction in the scatter of $M_\star$, $M_{\rm gas}$ and $Z_{\rm *}$ predictions, particularly at lower values. Additionally, our GNN reduces the bias of color predictions for the green valley subhalos. However, the improvement in other baryonic features is not substantial. The bottom part of each panel in Figure \ref{fig:allfeat} shows the number of subhalos in the predicted and true baryonic feature bins. Generally, the number distributions predicted by GNN and MLP are consistent with the true distribution for $M_\star$, $M_{\rm gas}$, $Z_{\rm *}$ and SFR, while those of color and $Z_{\rm gas}$ are less consistent. 

To further investigate the scatter in $M_\star$, we show the stellar mass growth history in the main progenitor branches of the lower-mass [$\log(M_\star/M_\odot)=10.0-10.5$] and higher-mass [$\log(M_\star/M_\odot)=10.5-11.5$] galaxy populations in Figure~\ref{fig:masshist}a. The predicted and the true mass history exhibit excellent consistency in both populations up to  $z=5$. Figure \ref{fig:masshist}b and Figure \ref{fig:masshist}c show the errors in the stellar mass growth history of the two populations. These panels indicate that there is a small bias in the stellar mass growth history for both populations. While the bias of the lower mass population does not show dependence on redshift, that of the higher mass population becomes slightly larger at low-$z$. Additionally, the scatter for higher and lower mass populations increases slightly at high-$z$. 

\section{Discussion and Prospects}
\label{sec:fin}
\subsection{Assessing and Understanding the Performance of our Model}
Although our model achieves an accurate mapping between the dark matter halo merger history and some relevant baryonic properties of galaxies such as $M_\star$, $Z_\star$, and $M_{\rm gas}$, the predictions of SFR, color, and $Z_{\rm gas}$ still show a relatively large scatter compared to the true values. This can be attributed to the stochastic nature of the star formation rate, gas metallicity, and color history. As shown in Table \ref{tab:comp}, these targets also have large chaotic uncertainties, and consistently show very little improvement when including merger history, since the large stochasticity implies that past information is not very informative.

\begin{figure*}[hbt!]
  \centering
  \includegraphics[width=1.0\linewidth]{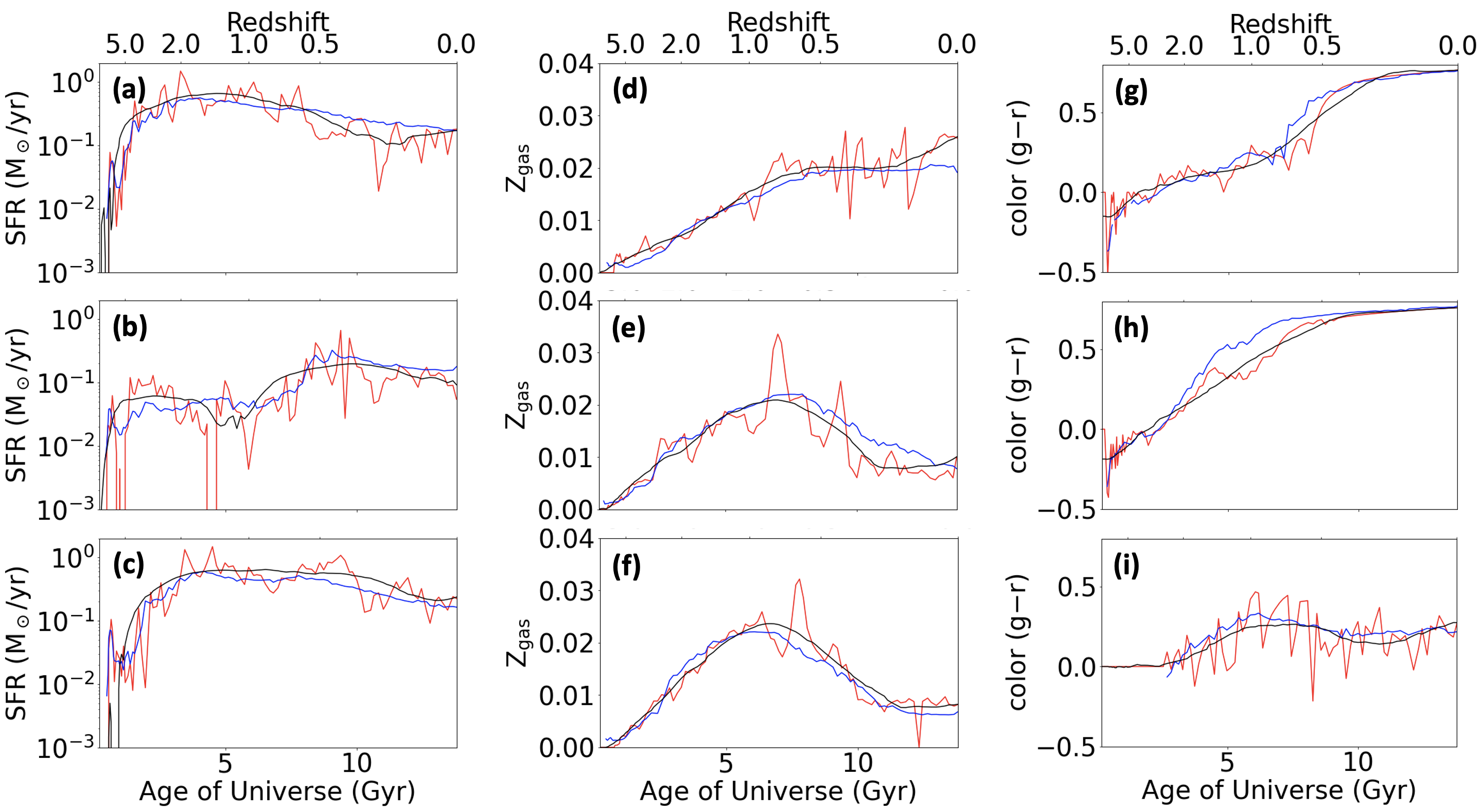}
  \caption{The predicted (blue), true (red), and smoothed (black) value of star formation history (panels a, b, and c), gas metallicity evolution (panels d, e, and f), and color history (panels g, h and i) from the main primary branch of each merger tree. A Savitzky–Golay filter with a window size of 27 snapshots and an order of 3 is applied to the red curves to generate the black curves, which represent the overall behavior of the history of different stochastic features. We randomly choose a subhalo for each panel to plot their evolution. The panels demonstrate the ability of our GNN model to predict the overall behavior of stochastic baryonic features. However, it struggles to recover the stochastic variations, which might be caused by the {\it IllustrisTNG} implementation of the star formation mechanism and other feedback mechanisms, as well as the resolution effects. While the overall history of these properties is vital to understanding galaxy evolution, the actual role stochastic processes play in the Universe is hard to pin down as our model mainly relies on the subgrid physics implemented in the TNG simulations.}
  \label{fig:stoch}
\end{figure*}

Figure \ref{fig:stoch} shows the predicted and true histories of star formation, gas metallicity, and color for a few randomly selected subhalos. We have picked three subhalos for each feature, resulting in 9 different subhalos. The panels demonstrate that the GNN can generally capture the overall trends of SFR, color, and gas metallicity histories, but it struggles to recover the stochastic variations and therefore performs similarly to models not including merger histories. However, as discussed by \citet{genel2019}, part of this stochasticity is due to the ways star formation and feedback are implemented in {\it IllustrisTNG}, as well as the resolution effects, and thus, the role analogous processes play in nature is hard to pin down. Therefore, even an \textit{ideal machine} would struggle to reproduce these quantities to a significantly higher fidelity. Regarding the color, the GNN accurately reproduces the subhalos in the red and blue populations; however, it struggles to predict exactly when the transition from the blue cloud to the red sequence occurs. This is consistent with the findings shown in Figure \ref{fig:allfeat}b, namely the scatter of subhalos in the green valley is larger compared to those in the red sequence and blue cloud. Nonetheless, for the population without a significant transition from the blue cloud to the red sequence, the stochastic variation is the main obstacle for GNN to predict the color history. 

We investigate our assumption by assessing the model's performance under abrupt changes in SFR by measuring the derivative of the star formation history (SFH) in the main progenitor branch of the merger trees, similar to the method outlined in \citet{Chuang2023}. We divide the data into two groups based on the \textit{magnitude} of the derivative of SFH. Specifically, we use a threshold of $10\,M_\odot\,$yr$^{-1}$\,Gyr$^{-1}$ to distinguish between a group with significant variations and another with smoother variations. Subsequently, we calculate the scatter of the SFR predictions for each group. The scatter and bias for the group with drastic variations are 0.38 and 0.19, respectively, whereas those for the smoother group are 0.33 and 0.001, respectively. These results show that our model performs less accurately when faced with a highly stochastic target. 

To further characterize the influence of the stochastic components on different baryonic features, we extract the baryonic feature history along the main primary branch of each merger tree and apply a Savitzky–Golay filter \citet[][]{savitzky1964} to remove the stochastic component. Then, we gauge the ability of the GNN model to predict the smoothed and unsmoothed baryonic feature history with scatter $\sigma$, $\rho$, and $R^2$ as presented in Table \ref{tab:compsmth}. The smoothed baryonic feature history (black curves) in Figure \ref{fig:stoch} generally follows the unsmoothed (true) baryonic feature history (red curves). For the color evolution, although the smoothed curve cannot represent the abrupt transition from the blue cloud to the red sequence, it can still trace the overall color history of the subhalos without the transition. For all features, the GNN traces the smoothed history better than that of the stochastic components, indicating the ability of our GNN model to predict the overall behavior of baryonic features while struggling to recover the stochastic variations. Nonetheless, the influences of the stochastic features on baryonic features are different. We demonstrate this point by calculating the relative difference in the scatter between the smoothed history and the GNN prediction  based on the unsmoothed history. The difference is larger for SFR, color, and $Z_{\rm gas}$ than for $M_\star$, $Z_\star$, and $M_{\rm gas}$, which indicates a larger influence of stochastic variation on SFR, color, and $Z_{\rm gas}$. 

\begin{deluxetable*}{ll|lllllll}
\tablecaption{Comparison of the GNN prediction on the true and smoothed baryonic feature history with three of the metrics. The metrics subscripted with ``Main'' and ``Smooth'' are the metrics calculated with the baryonic feature history from the main primary branch and the smoothed baryonic feature history, respectively. A Savitzky–Golay filter with a window size of 27 snapshots and an order of 3 is applied to the baryonic feature history to remove the stochastic component and generate the smoothed baryonic feature history. The best performance is shown in bold for each metric. For all features, GNN traces the smoothed history better than that with the stochastic components, indicating the ability of our GNN model to predict the overall behavior of baryonic features while struggling to recover the stochastic variations. Generally, for all baryonic features, the larger improvement from $\sigma_{\rm Main}$ to $\sigma_{\rm Smooth}$ would indicate larger stochasticity in the baryonic feature history. 
\label{tab:compsmth}}
\tablehead{
Target& $\sigma$&$\sigma_{\rm Main}$& $\sigma_{\rm Smooth}$&Difference\tablenotemark{a}&$\rho_{\rm Main}$& $\rho_{\rm Smooth}$&$R^2_{\rm Main}$& $R^2_{\rm Smooth}$\\
\hline
\multicolumn{9}{c}{redshift $0-2$}}
\startdata
\hline
$M_*$& 0.145& \textbf{0.144} & \textbf{0.144} & 0.0\% & \textbf{0.979} & \textbf{0.979} & \textbf{0.959} & \textbf{0.959}\\
\hline
SFR& 0.331 & 0.329 & \textbf{0.26 }& 21.1\% & 0.88 & \textbf{0.92 }& 0.774 & \textbf{0.846}\\
\hline
color ($g-r$)&0.108 &0.102 & \textbf{0.07 }& 31.1\% & 0.792 & \textbf{0.888} & 0.626 & \textbf{0.788}\\
\hline
$M_{\rm gas}$& 0.127 & 0.124 & \textbf{0.121} & 2.9\% & 0.946 & \textbf{0.948} & 0.894 & \textbf{0.899}\\
\hline
$Z_{\rm gas}$&0.158&0.159 & \textbf{0.143} & 9.8\% & 0.866 & \textbf{0.883} & 0.75 & \textbf{0.779}\\
\hline
$Z_{\rm \star}$& 0.125& 0.127 & \textbf{0.116} & 8.8\% & 0.926 & \textbf{0.935} & 0.857 & \textbf{0.875}
\enddata
\tablenotetext{a}{Defined as $(\sigma_{\rm MPB}-\sigma_{\rm smth})/\sigma_{\rm smth}$.}
\end{deluxetable*}

The higher accuracy of the predictions for $M_\star$, $M_{\rm gas}$, and $Z_\star$ may imply that, the merger history of a galaxy is more important for determining $M_\star$, $M_{\rm gas}$, and $Z_\star$ than for the other properties investigated in this work. However, as discussed above, this could also mean that we do not yet have enough simulated data to accurately learn the behaviors of a very stochastic target.

As the model continually improves with more data, the next generation of large hydrodynamical simulations will surely be able to improve the ability to accurately learn mappings between merger trees and baryons \citep{pakmor2022}. There is also the possibility of improving our model with a GNN taking into account the environmental information like the one presented by \cite{WuJespersen23}. Improving the method we rearrange the merger trees could also improve our model. In Section \ref{sec:gnn}, we assume that merger events are significant in the growth of dark matter subhalos, and therefore, we remove some of the original connections among subhalos in the merger trees that correspond to the smooth accretion mode of the subhalos' growth. Nonetheless, because the smooth accretion mode could also affect the evolution of baryonic features, leaving the original connections that represent the smooth accretion mode in the merger trees could also improve the performance of the GNN model.

\subsection{Potential Applications of the Model}
Despite the poorer performance for features with intrinsically large stochasticity, by leveraging the merger history, our model outperforms the state-of-the-art tools for learning the baryonic properties of galaxies and mapping these along the dark matter halo merger trees for all properties. The model works over a wide range of redshifts and works on \textit{every} branch of all merger trees. Utilizing the model, it is possible to ``emulate'' the result of MHD simulations if one has a  DMO simulation. One application of this is to recover the color-dependent 2PCFs in DMO simulations, given the color prediction from GNN. This provides an alternative way to test various AM schemes by conducting AM separately on blue and red subhalos and see whether the AM scheme can reproduce the color-dependent 2PCFs, such as the test performed in  \citet{Chuang2023}. 

Another potential application of our model is to apply it to the dark matter subhalos within constrained simulations. Presently, constrained simulations such as those conducted by \citet{wang2016} and \citet{mcalpine2022}, primarily employ DMO simulations, wherein the initial conditions are optimized to replicate the dark matter density fields of the local universe as inferred from observed galaxy distribution. After the optimization, each subhalo in the constrained simulation has one corresponding galaxy in the true universe. With our GNN model serving as an efficient and suitably accurate emulator, it becomes possible to swiftly incorporate baryonic properties into the DMO simulations. This offers the following benefits:
\begin{itemize}
    \item Previously, real galaxies and the subhalos in the MHD simulations have been compared only through statistical properties of certain populations of galaxies and subhalos, such as the fundamental plane \citep[][]{lu2020}, the star-forming main sequence \citep[][]{speagle2014, donnari2019}, etc. Nonetheless, by comparing the baryonic properties of galaxies estimated by the GNN in the DMO subhalos with the properties of true galaxies, it becomes feasible to establish a one-to-one comparison between the subhalos incorporated with TNG physics learned from the GNN and actual galaxies. 
    \item With the approach outlined in the previous bullet point, we can further compare the GNN-predicted SFH with those derived from spectral energy distribution (SED) fitting in order to calibrate existing SED fitting models, such as the one presented in  \citet{abdurrouf2021}. 
    \item By training the GNN with the merger trees in the constrained simulations and the baryonic properties of true galaxies, it is possible to learn the relation between merger trees and galaxies in the universe. 
    \item Assuming that the last step can be done, it becomes possible to directly optimize the initial conditions of DMO simulations to match the distribution of galaxies instead of relying on the inferred  density field. This is achievable because we can estimate whether a subhalo contains galaxies and what baryonic properties the galaxy residing in the subhalo has.
    
\end{itemize}

\begin{acknowledgments}
We acknowledge support from the National Science and Technology Council of Taiwan under grants MOST 110-2112-M-001-004, MOST 111-2112-M-001-043, and NSTC 112-2112-M-001-061. We thank Dylan Nelson for helpful comments on IllustrisTNG simulation suite. CKJ thanks Zachary S.~Hemler, John F.~Wu and Risa H.~Wechsler for useful comments. YTL thanks IH, LYL and ALL for constant encouragement and inspiration.

The numerical work was conducted on the high-performance computing facility at the Institute of Astronomy and Astrophysics in Academia Sinica (https://hpc.tiara.sinica.edu.tw). The IllustrisTNG simulations were undertaken with compute time awarded by the Gauss Centre for Supercomputing (GCS) under GCS Large-Scale Projects GCS-ILLU and GCS-DWAR on the GCS share of the supercomputer Hazel Hen at the High Performance Computing Center Stuttgart (HLRS), as well as on the machines of the Max Planck Computing and Data Facility (MPCDF) in Garching, Germany.

\end{acknowledgments}

\clearpage
\newpage

\appendix
\section{Dark Matter Subhalo Features Used}
\label{sec:fea}
\centerline{\textbf{Subhalo/Halo Features in Nodes}}
\begin{itemize}
    \item $M_{\rm sub}$: Total mass of all member particles which are bound to this Subhalo. Particles bound to subhaloes of this Subhalo are not accounted for. [in $M_\odot$]
    \item $R_{\rm h}$: Comoving radius containing half of the $M_{\rm DM,sub}$. [in ckpc]
    \item $V_{\rm max}$: Maximum value of the subhalo spherically-averaged rotation curve at its redshift [in $\rm{km\ s^{-1}}$]
    \item $R_{\rm max}$: Comoving radius of rotation curve maximum (where $V_{\rm max}$ is achieved) [in ckpc]
    \item $M_{\rm max}$: Subhalo mass within $R_{\rm max}$ [in $M_\odot$]
    \item $V_{\rm disp}$: One-dimensional velocity dispersion of all the member particles (the 3D dispersion devided by $\sqrt{3}$). [in $\rm{km\ s^{-1}}$]
    \item $J$: Total spin, computed for each subhalo as the mass weighted sum of the relative coordinate times relative velocity of all member particles. [in $\rm{kpc\ km\ s^{-1}}$]
    \item $\delta_{\rm pos,CM}$: Distance between the spatial position of the particle with the minium gravitational potential energy ($X_{\rm pos,sub}$) and the center of mass ($X_{\rm CM,sub}$) of the subhalo (the sum of the mass weighted relative coordinates of all particles in the subhalo) [in ckpc]
    \item $\delta X_{\rm sub,halo}$: Distance between $X_{\rm pos,sub}$ and the spatial position of the particle with the minimum gravitational potential energy in the FoF halo ($X_{\rm pos,halo}$) the subhalo belongs to. [in ckpc]
    \item $\delta V_{\rm subhalo,halo}$: Absolute velocity relative to the velocity of the FoF halo, computed as the sum of the mass weighted velocities of all particles in the subhalo. [in $\rm{km\ s^{-1]}}$]
    \item $\delta X_{\rm subhalo, MP}$: Distance between $X_{\rm pos}$ of the subhalo and the $X_{\rm pos}$ of the subhalo in the main primary branch at the same snapshot. [in ckpc]
    \item $M_{\rm halo,200m}$: Total Mass of this FoF halo (which the subhalo is located) enclosed in a sphere whose mean density is 200 times the mean density of the universe, at the time the halo is considered. [in $M_\odot$]
    \item $R_{\rm halo,200m}$: Comoving Radius of a sphere centered at the $X_{\rm pos,halo}$ of the FoF halo whose mean density is 200 times the mean density of the Universe, at the time the halo is considered. [in ckpc]
    \item $t_{\rm age}$: Age of the universe at the time subhalo is considered. [in Gyr]
    \item cent or sat?: Whether the subhalo is a central (1) or satellite (0) in the FoF halo. [binary format]
\end{itemize}
\centerline{\textbf{Subhalo/Halo Features in Edges}}
The difference of the node features between every pair of subhalos linked by an edge in the merger tree.
\begin{itemize}
    \item $\Delta X_{\rm pos}$: Distance of $X_{\rm pos}$ of the two subhalos [in ckpc]
    \item Difference between $\delta X_{\rm subhalo, MP}$ of the two subhalos. [in ckpc]
    \item $\Delta V_{\rm subhalo,halo}$: Difference between $\delta V_{\rm subhalo,halo}$ of the two subhalos [in $\rm{km\ s^{-1}}$]
    \item Difference between $J$ of the two subhalos [in $\rm{kpc\ km\ s^{-1}}$]
\end{itemize}

\centerline{\textbf{Global Features of the Merger Trees}}
\begin{itemize}
    \item $N_{\rm prog}$: Number of progenitors for the merger tree.
    \item $N_{\rm merge}$: Number of merger events for the merger tree.
\end{itemize}

\section{Hyper-Parameter Search}
\label{sec:hyp}
The hyper-parameters we searched for GNN include the number of MLP layers \{2, \textbf{3}\} to decode the node features from the GNN, the number of \texttt{MetaLayers} \{2, 3, \textbf{4}\}, different aggregation methods (sum or mean aggregation), the initial learning rate \{0.01, \textbf{0.001}\} for the {\it OneCycleLR} policy \citep{smith17}, the relative strengths of classification loss with respect to the regression loss ($h$ in Equation~\ref{eq:tot}, \{\textbf{1.0}, 100.0\}), and the relative strengths of L1 and L2 norm of all parameters in the GNN with respect to the regression loss ($l_1$ and $l_2$ in Equation~\ref{eq:tot}, \{\textbf{0.0}, 0.0001\}). The best combination of the hyper-parameters is shown in bold. Due to the limitation of the computing resources, we are only able to search a small fraction of the parameter space for each hyper-parameter. We welcome others to test our GNN model in more detail. The specifics of the computational resources  used are listed below:
\begin{enumerate}
    \item GPU number: 6 NVIDIA Quadro RTX 8000 GPUs
    \item GPU memory: 48 GB/GPU
    \item Number of train epochs: 2000 per hyper-parameter search
    \item Training time: $\sim$7 hrs per hyper-parameter search
    \item Total hyper-parameter training time: $\sim18$ days
\end{enumerate}

\section{Uncertainties}
\label{sec:uncert}
\begin{figure*}
  \centering
  \includegraphics[width=1.0\linewidth]{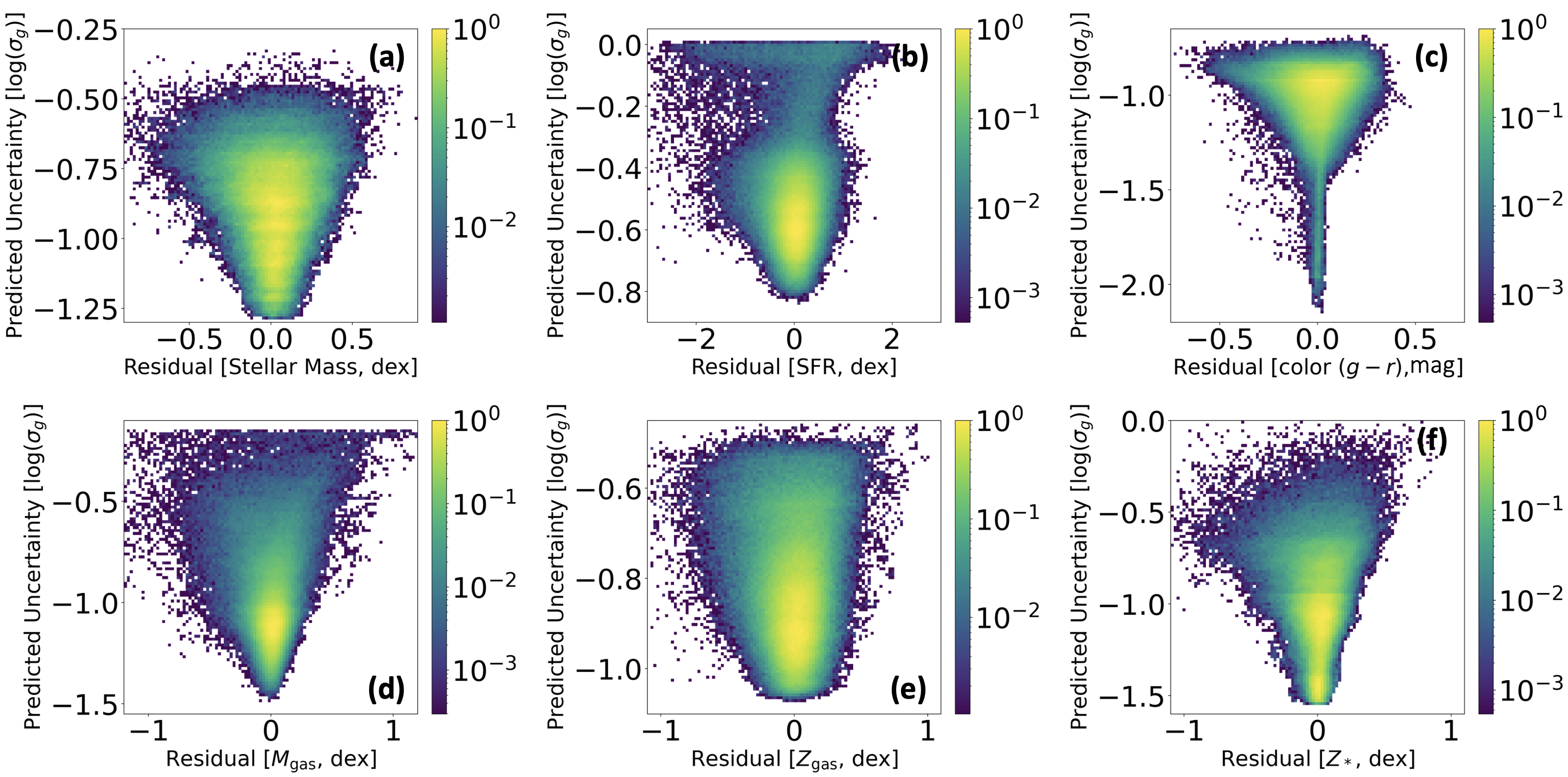}
  \caption{Each panel shows the predicted uncertainty and the residual between the prediction and the true value for a specific baryonic feature  (indicated at the bottom of each panel).
  The data point densities are normalized such that the maximum value is one. The $y$-axis in each panel is logarithmically scaled. In general, the model successfully assigns large uncertainties to targets that are likely to have large residuals. Although the data point densities exhibit an overall axial symmetry along the axis of $\Delta y_i=0$, asymmetric features appear with higher predicted uncertainties for SFR, color, M$_{\rm gas}$, and $Z_\star$. The asymmetric features could be due to the overestimation of values with high stochasticities, which results in higher predicted uncertainty. }
  \label{fig:uncert}
\end{figure*}

Figure \ref{fig:uncert} shows the predicted uncertainty ($\widehat{\sigma}$ in Equation~\ref{eq:gauss}) and the residual between the prediction and the true value ($\Delta y_i$ in Equation~\ref{eq:scat}) for each baryonic feature. In general, the scatter of all predicted features tends to increase with higher predicted uncertainty, which indicates that the predicted Gaussian uncertainties faithfully reflect the model's ability to predict a given target for a given merger tree. Although the data point densities exhibit an overall axial symmetry along the axis of $\Delta y_i=0$, asymmetric features appear with higher predicted uncertainties for SFR, color, M$_{\rm gas}$, and $Z_\star$. Another noticeable feature in Figure \ref{fig:uncert} is the bimodality in the uncertainties for SFR and $Z_\star$. 

The asymmetric features could be attributed to the overestimation of values with high stochasticities. As described in Section \ref{sec:fin}, we divide the data into two groups based on the \textit{magnitude} of the derivative of the SFH. We then measure the mean uncertainty of the SFR for the two groups. The mean uncertainty for the group with drastic variations is 1.89, whereas that for the smoother group is 1.43. These results show that our model predicts a higher uncertainty for targets with higher stochasticity, while Section \ref{sec:fin} indicates that bias is also more significant with higher stochasticity. Thus, the targets with higher uncertainty show a positive residual in Figure \ref{fig:uncert}b.

\begin{figure*}
  \centering
  \includegraphics[width=0.9\linewidth]{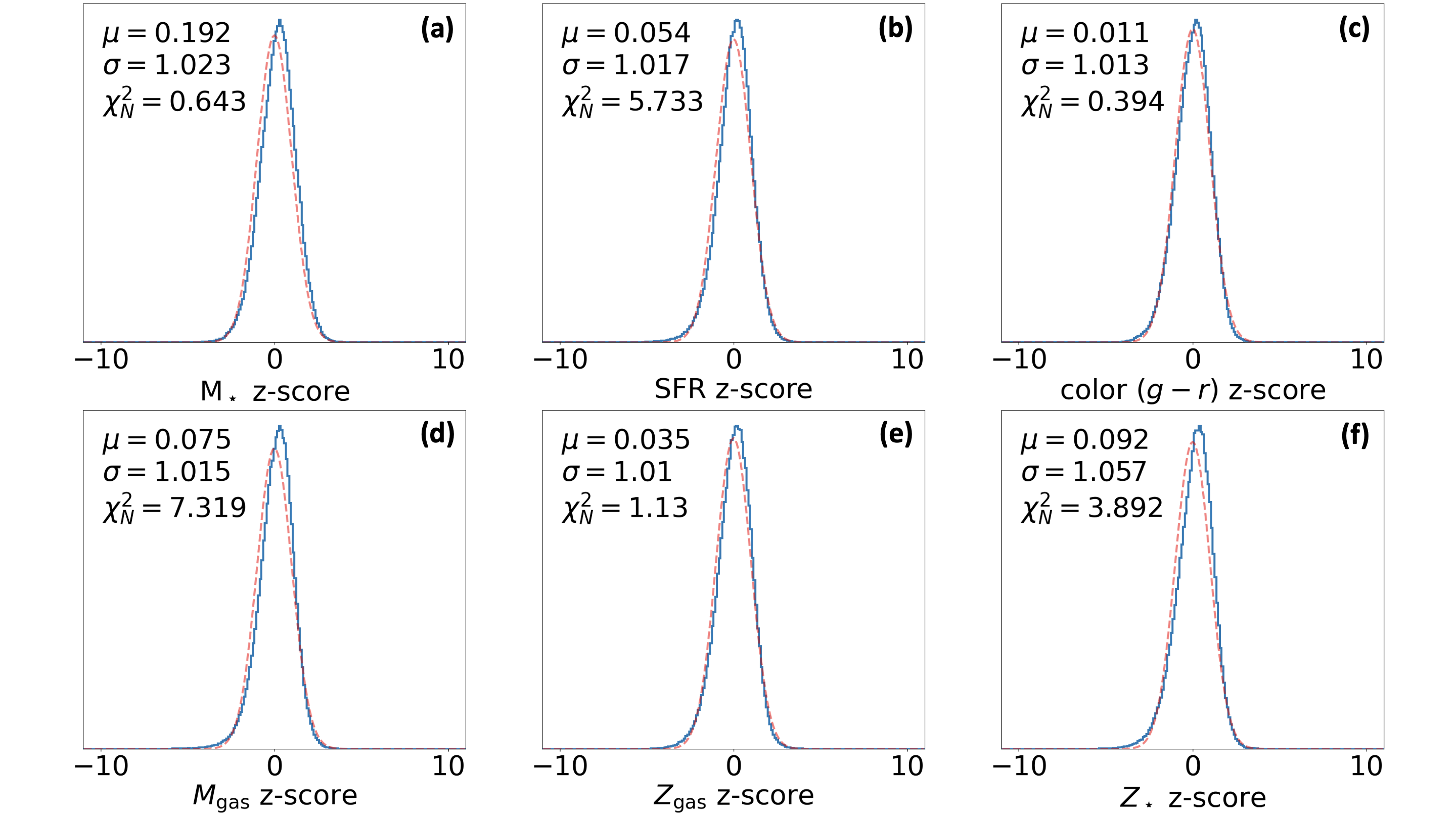}
  \caption{Each panel shows the distribution of pulls/$z$-scores 
  ($z$-score$_i=\Delta y_i/\widehat{\sigma_i}$, shown in blue),
  and a unit Gaussian (shown in red) for reference. The mean, variance, and $\chi^2_N$ for each feature are displayed in the upper-left of each panel. In general, although asymmetric features appear in Figure \ref{fig:uncert}, the overall distributions of $z$-scores can still be approximated by a Gaussian. }
  \label{fig:zscore}
\end{figure*}

Since the model is minimized with a Gaussian loss, we expect that the pull/$z$-score ($z\equiv \Delta y/\hat{\sigma}$) should show a Gaussian distribution, while the $\chi^2_N=\chi^2/N$ should be close to 1. Figure \ref{fig:zscore} shows the distribution of $z$-scores (the pull plot) and a unit Gaussian. Similar to \citet[][]{jespersen2022}, the $z$-score distributions in Figure \ref{fig:zscore} can be approximated by a unit Gaussian, indicating that error estimation is highly accurate. 

\clearpage
\newpage

\section{Full Table of Metrics}
\label{app:fulmet}
\begin{deluxetable*}{l|llllllllll}[!htb]
\tablecaption{The full table of metrics used to quantify the performance of GNN, MLP, and AM.\label{tab:comp1}}
\tablehead{
Target& Method& $\sigma$& bias & $\rho$& $R^2$& F1&TPR&FPR&FNR&TNR\\
\hline
\multicolumn{11}{c}{redshift 0}}
\startdata
$M_*$& GNN& 0.141& 0.02& 0.979& 0.959& 1.0 & 99.95& 0.0& 0.05& 0.0\\
& MLP& 0.187& -0.009& 0.963& 0.928& 1.0 &100.0& 0.0& 0.0& 0.0\\
& AM$_{\psi_5}$ \tablenotemark{b}& 0.184 & 0.0&0.959& 0.920\\
& AM$_{V_{\rm 90\%}}$ \tablenotemark{c}&0.222& 0.0 & 0.963 & 0.927\\
\hline
SFR& GNN& 0.388& 0.014& 0.767& 0.589& 0.968 & 85.91& 3.15& 2.6& 8.34\\
&MLP& 0.4& 0.009& 0.75& 0.562& 0.964& 85.93& 3.87& 2.59& 7.61\\
\hline
color ($g-r$)&GNN& 0.129& -0.001& 0.693& 0.48& 1.0 & 99.99& 0.0& 0.01& 0.0\\
& MLP& 0.133& -0.012& 0.667& 0.444& 1.0 & 100.0& 0.0& 0.0& 0.0\\
\hline
$M_{\rm gas}$&GNN&0.187& 0.013& 0.88& 0.774& 0.994& 97.82& 0.7& 0.48& 1.0\\
& MLP& 0.216& 0.002& 0.844& 0.713& 0.994& 98.19& 1.14& 0.11& 0.56\\
\hline
$Z_{\rm gas}$&GNN&0.179& 0.005& 0.773& 0.598& 97.1& 1.2& 0.53& 1.17\\
& MLP& 0.186& -0.016& 0.749& 0.562& 0.99& 97.47& 1.79& 0.16& 0.57\\
\hline
$Z_*$& GNN& 0.111& 0.005& 0.918& 0.843& 1.0& 99.97& 0.0& 0.03& 0.0\\
& MLP& 0.123& -0.005& 0.9& 0.81& 1.0& 100.0& 0.0& 0.0& 0.0\\
\hline
\hline
\multicolumn{11}{c}{redshift $0-2$}\\
\hline
$M_*$& GNN& 0.145& 0.022& 0.979& 0.958& 0.999& 99.59& 0.23& 0.06& 0.11\\
& MLP& 0.184& -0.011& 0.966& 0.934& 0.998& 99.6& 0.26& 0.05& 0.1\\
\hline
SFR& GNN& 0.331& 0.014& 0.872& 0.76& 0.985& 94.34& 1.78& 1.03& 2.86\\
& MLP& 0.348& 0.005& 0.859& 0.738& 0.983& 94.1& 2.04& 1.18& 2.68\\
\hline
color ($g-r$)& GNN& 0.108& 0.001& 0.777& 0.603& 0.999& 99.59& 0.23& 0.06& 0.12\\
& MLP& 0.112& -0.006& 0.76& 0.577& 0.998& 99.59& 0.26& 0.05& 0.1\\
\hline
$M_{\rm gas}$&GNN& 0.126& 0.007& 0.944& 0.891& 0.998& 99.18& 0.35& 0.13& 0.35\\
& MLP& 0.143& 0.005& 0.928& 0.862& 0.997& 99.14& 0.43& 0.14& 0.29\\
\hline
$Z_{\rm gas}$& GNN& 0.158& 0.004& 0.864& 0.746& 0.996& 98.85& 0.62& 0.14& 0.39\\
& MLP& 0.17& -0.005& 0.846& 0.715& 0.996& 98.78& 0.71& 0.18& 0.33\\
\hline
$Z_*$& GNN& 0.128& 0.009& 0.923& 0.852& 0.999& 99.57& 0.26& 0.06& 0.11\\
& MLP& 0.144& -0.005& 0.905& 0.818& 0.998& 99.57& 0.29& 0.04& 0.09\\
\hline
\enddata
\tablenotetext{b}{Please refer to Table \ref{tab:comp} for the definition. }
\tablenotetext{c}{Please refer to Table \ref{tab:comp} for the definition. }
\end{deluxetable*}
\clearpage
\newpage

\bibliography{bib}

\begin{thebibliography}{}
\expandafter\ifx\csname natexlab\endcsname\relax\def\natexlab#1{#1}\fi
\providecommand{\url}[1]{\href{#1}{#1}}
\providecommand{\dodoi}[1]{doi:~\href{http://doi.org/#1}{\nolinkurl{#1}}}
\providecommand{\doeprint}[1]{\href{http://ascl.net/#1}{\nolinkurl{http://ascl.net/#1}}}
\providecommand{\doarXiv}[1]{\href{https://arxiv.org/abs/#1}{\nolinkurl{https://arxiv.org/abs/#1}}}

\bibitem[{{Abdurro'uf} {et~al.}(2021){Abdurro'uf}, {Lin}, {Wu}, \& {Akiyama}}]{abdurrouf2021}
{Abdurro'uf}, {Lin}, Y.-T., {Wu}, P.-F., \& {Akiyama}, M. 2021, \apjs, 254, 15, \dodoi{10.3847/1538-4365/abebe2}

\bibitem[{{Agarwal} {et~al.}(2018){Agarwal}, {Dav{\'e}}, \& {Bassett}}]{shankar2018}
{Agarwal}, S., {Dav{\'e}}, R., \& {Bassett}, B.~A. 2018, \mnras, 478, 3410, \dodoi{10.1093/mnras/sty1169}

\bibitem[{{Battaglia} {et~al.}(2018){Battaglia}, {Hamrick}, {Bapst}, {Sanchez-Gonzalez}, {Zambaldi}, {Malinowski}, {Tacchetti}, {Raposo}, {Santoro}, {Faulkner}, {Gulcehre}, {Song}, {Ballard}, {Gilmer}, {Dahl}, {Vaswani}, {Allen}, {Nash}, {Langston}, {Dyer}, {Heess}, {Wierstra}, {Kohli}, {Botvinick}, {Vinyals}, {Li}, \& {Pascanu}}]{battaglia2018}
{Battaglia}, P.~W., {Hamrick}, J.~B., {Bapst}, V., {et~al.} 2018, arXiv e-prints, arXiv:1806.01261.
\newblock \doarXiv{1806.01261}

\bibitem[{{Breivik} {et~al.}(2022){Breivik}, {Connolly}, {Ford}, {Juri{\'c}}, {Mandelbaum}, {Miller}, {Norman}, {Olsen}, {O'Mullane}, {Price-Whelan}, {Sacco}, {Sokoloski}, {Villar}, {Acquaviva}, {Ahumada}, {AlSayyad}, {Alves}, {Andreoni}, {Anguita}, {Best}, {Bianco}, {Bonito}, {Bradshaw}, {Burke}, {Rodrigues de Campos}, {Cantiello}, {Caplar}, {Chandler}, {Chan}, {Nicolaci da Costa}, {Danieli}, {Davenport}, {Fabbian}, {Fagin}, {Gagliano}, {Gall}, {Garavito Camargo}, {Gawiser}, {Gezari}, {Gomboc}, {Gonzalez-Morales}, {Graham}, {Gschwend}, {Guy}, {Holman}, {Hsieh}, {Hundertmark}, {Ili{\'c}}, {Ishida}, {Jurki{\'c}}, {Kannawadi}, {Kosakowski}, {Kova{\v{c}}evi{\'c}}, {Kubica}, {Lanusse}, {Lazar}, {Levine}, {Li}, {Lu}, {Luna}, {Mahabal}, {Malz}, {Mao}, {Medan}, {Moeyens}, {Nikoli{\'c}}, {Nikutta}, {O'Dowd}, {Olsen}, {Pearson}, {Villicana Pedraza}, {Popinchalk}, {Popovi{\'c}}, {Pritchard}, {Quint}, {Radovi{\'c}}, {Ragosta}, {Riccio}, {Riley}, {Ro{\.z}ek}, {S{\'a}nchez-S{\'a}ez}, {Sarro}, {Saunders}, {Savi{\'c}},
  {Schmidt}, {Scott}, {Shirley}, {Smotherman}, {Stetzler}, {Storey-Fisher}, {Street}, {Trilling}, {Tsapras}, {Ustamujic}, {van Velzen}, {V{\'a}zquez-Mata}, {Venuti}, {Wyatt}, {Yu}, \& {Zabludoff}}]{lsst22}
{Breivik}, K., {Connolly}, A.~J., {Ford}, K.~E.~S., {et~al.} 2022, arXiv e-prints, arXiv:2208.02781.
\newblock \doarXiv{2208.02781}

\bibitem[{{Chuang} \& {Lin}(2023)}]{Chuang2023}
{Chuang}, C.-Y., \& {Lin}, Y.-T. 2023, \apj, 944, 207, \dodoi{10.3847/1538-4357/acb5f3}

\bibitem[{{Croton} {et~al.}(2016){Croton}, {Stevens}, {Tonini}, {Garel}, {Bernyk}, {Bibiano}, {Hodkinson}, {Mutch}, {Poole}, \& {Shattow}}]{croton2016}
{Croton}, D.~J., {Stevens}, A. R.~H., {Tonini}, C., {et~al.} 2016, \apjs, 222, 22, \dodoi{10.3847/0067-0049/222/2/22}

\bibitem[{{de Santi} {et~al.}(2022){de Santi}, {Rodrigues}, {Montero-Dorta}, {Abramo}, {Tucci}, \& {Artale}}]{natali2022}
{de Santi}, N. S.~M., {Rodrigues}, N. V.~N., {Montero-Dorta}, A.~D., {et~al.} 2022, \mnras, 514, 2463, \dodoi{10.1093/mnras/stac1469}

\bibitem[{{Donnari} {et~al.}(2019){Donnari}, {Pillepich}, {Nelson}, {Vogelsberger}, {Genel}, {Weinberger}, {Marinacci}, {Springel}, \& {Hernquist}}]{donnari2019}
{Donnari}, M., {Pillepich}, A., {Nelson}, D., {et~al.} 2019, \mnras, 485, 4817, \dodoi{10.1093/mnras/stz712}

\bibitem[{{Genel} {et~al.}(2019){Genel}, {Bryan}, {Springel}, {Hernquist}, {Nelson}, {Pillepich}, {Weinberger}, {Pakmor}, {Marinacci}, \& {Vogelsberger}}]{genel2019}
{Genel}, S., {Bryan}, G.~L., {Springel}, V., {et~al.} 2019, \apj, 871, 21, \dodoi{10.3847/1538-4357/aaf4bb}

\bibitem[{Good(2018)}]{good1952}
Good, I.~J. 2018, Journal of the Royal Statistical Society: Series B (Methodological), 14, 107, \dodoi{10.1111/j.2517-6161.1952.tb00104.x}

\bibitem[{{Jespersen} {et~al.}(2022){Jespersen}, {Cranmer}, {Melchior}, {Ho}, {Somerville}, \& {Gabrielpillai}}]{jespersen2022}
{Jespersen}, C.~K., {Cranmer}, M., {Melchior}, P., {et~al.} 2022, \apj, 941, 7, \dodoi{10.3847/1538-4357/ac9b18}

\bibitem[{{Kamdar} {et~al.}(2016){Kamdar}, {Turk}, \& {Brunner}}]{harshil2016}
{Kamdar}, H.~M., {Turk}, M.~J., \& {Brunner}, R.~J. 2016, \mnras, 455, 642, \dodoi{10.1093/mnras/stv2310}

\bibitem[{{Kauffmann} {et~al.}(1993){Kauffmann}, {White}, \& {Guiderdoni}}]{kauffmann1993}
{Kauffmann}, G., {White}, S.~D.~M., \& {Guiderdoni}, B. 1993, \mnras, 264, 201, \dodoi{10.1093/mnras/264.1.201}

\bibitem[{{Kravtsov} {et~al.}(2004){Kravtsov}, {Berlind}, {Wechsler}, {Klypin}, {Gottl{\" o}ber}, {Allgood}, \& {Primack}}]{kravtsov04}
{Kravtsov}, A.~V., {Berlind}, A.~A., {Wechsler}, R.~H., {et~al.} 2004, \apj, 609, 35

\bibitem[{{Lovell} {et~al.}(2022){Lovell}, {Wilkins}, {Thomas}, {Schaller}, {Baugh}, {Fabbian}, \& {Bah{\'e}}}]{christopher2022}
{Lovell}, C.~C., {Wilkins}, S.~M., {Thomas}, P.~A., {et~al.} 2022, \mnras, 509, 5046, \dodoi{10.1093/mnras/stab3221}

\bibitem[{{Lu} {et~al.}(2020){Lu}, {Xu}, {Wang}, {Mao}, {Ge}, {Springel}, {Wang}, {Vogelsberger}, {Naiman}, \& {Hernquist}}]{lu2020}
{Lu}, S., {Xu}, D., {Wang}, Y., {et~al.} 2020, \mnras, 492, 5930, \dodoi{10.1093/mnras/staa173}

\bibitem[{{Marinacci} {et~al.}(2018){Marinacci}, {Vogelsberger}, {Pakmor}, {Torrey}, {Springel}, {Hernquist}, {Nelson}, {Weinberger}, {Pillepich}, {Naiman}, \& {Genel}}]{Marinacci2018}
{Marinacci}, F., {Vogelsberger}, M., {Pakmor}, R., {et~al.} 2018, \mnras, 480, 5113, \dodoi{10.1093/mnras/sty2206}

\bibitem[{{McAlpine} {et~al.}(2022){McAlpine}, {Helly}, {Schaller}, {Sawala}, {Lavaux}, {Jasche}, {Frenk}, {Jenkins}, {Lucey}, \& {Johansson}}]{mcalpine2022}
{McAlpine}, S., {Helly}, J.~C., {Schaller}, M., {et~al.} 2022, \mnras, 512, 5823, \dodoi{10.1093/mnras/stac295}

\bibitem[{{Naab} \& {Ostriker}(2017)}]{naab17}
{Naab}, T., \& {Ostriker}, J.~P. 2017, \araa, 55, 59, \dodoi{10.1146/annurev-astro-081913-040019}

\bibitem[{{Naiman} {et~al.}(2018){Naiman}, {Pillepich}, {Springel}, {Ramirez-Ruiz}, {Torrey}, {Vogelsberger}, {Pakmor}, {Nelson}, {Marinacci}, {Hernquist}, {Weinberger}, \& {Genel}}]{Naiman2018}
{Naiman}, J.~P., {Pillepich}, A., {Springel}, V., {et~al.} 2018, \mnras, 477, 1206, \dodoi{10.1093/mnras/sty618}

\bibitem[{{Nelson} {et~al.}(2015){Nelson}, {Pillepich}, {Genel}, {Vogelsberger}, {Springel}, {Torrey}, {Rodriguez-Gomez}, {Sijacki}, {Snyder}, {Griffen}, {Marinacci}, {Blecha}, {Sales}, {Xu}, \& {Hernquist}}]{nelson2015}
{Nelson}, D., {Pillepich}, A., {Genel}, S., {et~al.} 2015, Astronomy and Computing, 13, 12, \dodoi{10.1016/j.ascom.2015.09.003}

\bibitem[{{Nelson} {et~al.}(2018){Nelson}, {Pillepich}, {Springel}, {Weinberger}, {Hernquist}, {Pakmor}, {Genel}, {Torrey}, {Vogelsberger}, {Kauffmann}, {Marinacci}, \& {Naiman}}]{Nelson18}
{Nelson}, D., {Pillepich}, A., {Springel}, V., {et~al.} 2018, \mnras, 475, 624, \dodoi{10.1093/mnras/stx3040}

\bibitem[{{Nelson} {et~al.}(2019){Nelson}, {Springel}, {Pillepich}, {Rodriguez-Gomez}, {Torrey}, {Genel}, {Vogelsberger}, {Pakmor}, {Marinacci}, {Weinberger}, {Kelley}, {Lovell}, {Diemer}, \& {Hernquist}}]{Nelson2019}
{Nelson}, D., {Springel}, V., {Pillepich}, A., {et~al.} 2019, Computational Astrophysics and Cosmology, 6, 2, \dodoi{10.1186/s40668-019-0028-x}

\bibitem[{{Oono} \& {Suzuki}(2019)}]{oono2019}
{Oono}, K., \& {Suzuki}, T. 2019, arXiv e-prints, arXiv:1905.10947.
\newblock \doarXiv{1905.10947}

\bibitem[{{Pakmor} {et~al.}(2022){Pakmor}, {Springel}, {Coles}, {Guillet}, {Pfrommer}, {Bose}, {Barrera}, {Delgado}, {Ferlito}, {Frenk}, {Hadzhiyska}, {Hern{\'a}ndez-Aguayo}, {Hernquist}, {Kannan}, \& {White}}]{pakmor2022}
{Pakmor}, R., {Springel}, V., {Coles}, J.~P., {et~al.} 2022, arXiv e-prints, arXiv:2210.10060, \dodoi{10.48550/arXiv.2210.10060}

\bibitem[{{Pillepich} {et~al.}(2018){Pillepich}, {Nelson}, {Hernquist}, {Springel}, {Pakmor}, {Torrey}, {Weinberger}, {Genel}, {Naiman}, {Marinacci}, \& {Vogelsberger}}]{pillepich2018}
{Pillepich}, A., {Nelson}, D., {Hernquist}, L., {et~al.} 2018, \mnras, 475, 648, \dodoi{10.1093/mnras/stx3112}

\bibitem[{{Planck Collaboration} {et~al.}(2016){Planck Collaboration}, {Ade}, {Aghanim}, {Arnaud}, {Ashdown}, {Aumont}, {Baccigalupi}, {Banday}, {Barreiro}, {Bartlett}, \& et~al.}]{planck16}
{Planck Collaboration}, {Ade}, P.~A.~R., {Aghanim}, N., {et~al.} 2016, \aap, 594, A13, \dodoi{10.1051/0004-6361/201525830}

\bibitem[{{Rodriguez-Gomez} {et~al.}(2015){Rodriguez-Gomez}, {Genel}, {Vogelsberger}, {Sijacki}, {Pillepich}, {Sales}, {Torrey}, {Snyder}, {Nelson}, {Springel}, {Ma}, \& {Hernquist}}]{gomez2015}
{Rodriguez-Gomez}, V., {Genel}, S., {Vogelsberger}, M., {et~al.} 2015, \mnras, 449, 49, \dodoi{10.1093/mnras/stv264}

\bibitem[{{Rossi} {et~al.}(2021){Rossi}, {Choi}, {Moon}, {Bautista}, {Gil-Mar{\'\i}n}, {Paviot}, {Vargas-Maga{\~n}a}, {de la Torre}, {Fromenteau}, {Ross}, {{\'A}vila}, {Burtin}, {Dawson}, {Escoffier}, {Habib}, {Heitmann}, {Hou}, {Mueller}, {Percival}, {Smith}, {Zhao}, \& {Zhao}}]{rossi21}
{Rossi}, G., {Choi}, P.~D., {Moon}, J., {et~al.} 2021, \mnras, 505, 377, \dodoi{10.1093/mnras/staa3955}

\bibitem[{{Savitzky} \& {Golay}(1964)}]{savitzky1964}
{Savitzky}, A., \& {Golay}, M.~J.~E. 1964, Analytical Chemistry, 36, 1627, \dodoi{10.1021/ac60214a047}

\bibitem[{{Smith} \& {Topin}(2017)}]{smith17}
{Smith}, L.~N., \& {Topin}, N. 2017, arXiv e-prints, arXiv:1708.07120.
\newblock \doarXiv{1708.07120}

\bibitem[{{Somerville} \& {Dav{\'e}}(2015)}]{somerville15}
{Somerville}, R.~S., \& {Dav{\'e}}, R. 2015, \araa, 53, 51, \dodoi{10.1146/annurev-astro-082812-140951}

\bibitem[{{Somerville} {et~al.}(2008){Somerville}, {Hopkins}, {Cox}, {Robertson}, \& {Hernquist}}]{somerville2008}
{Somerville}, R.~S., {Hopkins}, P.~F., {Cox}, T.~J., {Robertson}, B.~E., \& {Hernquist}, L. 2008, \mnras, 391, 481, \dodoi{10.1111/j.1365-2966.2008.13805.x}

\bibitem[{{Somerville} {et~al.}(2015){Somerville}, {Popping}, \& {Trager}}]{somerville2015}
{Somerville}, R.~S., {Popping}, G., \& {Trager}, S.~C. 2015, \mnras, 453, 4337, \dodoi{10.1093/mnras/stv1877}

\bibitem[{{Speagle} {et~al.}(2014){Speagle}, {Steinhardt}, {Capak}, \& {Silverman}}]{speagle2014}
{Speagle}, J.~S., {Steinhardt}, C.~L., {Capak}, P.~L., \& {Silverman}, J.~D. 2014, \apjs, 214, 15, \dodoi{10.1088/0067-0049/214/2/15}

\bibitem[{{Springel} {et~al.}(2001){Springel}, {White}, {Tormen}, \& {Kauffmann}}]{springel2001}
{Springel}, V., {White}, S. D.~M., {Tormen}, G., \& {Kauffmann}, G. 2001, \mnras, 328, 726, \dodoi{10.1046/j.1365-8711.2001.04912.x}

\bibitem[{{Springel} {et~al.}(2005){Springel}, {White}, {Jenkins}, {Frenk}, {Yoshida}, {Gao}, {Navarro}, {Thacker}, {Croton}, {Helly}, {Peacock}, {Cole}, {Thomas}, {Couchman}, {Evrard}, {Colberg}, \& {Pearce}}]{springel05}
{Springel}, V., {White}, S.~D.~M., {Jenkins}, A., {et~al.} 2005, \nat, 435, 629, \dodoi{10.1038/nature03597}

\bibitem[{{Springel} {et~al.}(2018){Springel}, {Pakmor}, {Pillepich}, {Weinberger}, {Nelson}, {Hernquist}, {Vogelsberger}, {Genel}, {Torrey}, {Marinacci}, \& {Naiman}}]{springel18}
{Springel}, V., {Pakmor}, R., {Pillepich}, A., {et~al.} 2018, \mnras, 475, 676, \dodoi{10.1093/mnras/stx3304}

\bibitem[{{Villaescusa-Navarro} {et~al.}(2020){Villaescusa-Navarro}, {Hahn}, {Massara}, {Banerjee}, {Delgado}, {Ramanah}, {Charnock}, {Giusarma}, {Li}, {Allys}, {Brochard}, {Uhlemann}, {Chiang}, {He}, {Pisani}, {Obuljen}, {Feng}, {Castorina}, {Contardo}, {Kreisch}, {Nicola}, {Alsing}, {Scoccimarro}, {Verde}, {Viel}, {Ho}, {Mallat}, {Wandelt}, \& {Spergel}}]{Villaescusa20}
{Villaescusa-Navarro}, F., {Hahn}, C., {Massara}, E., {et~al.} 2020, \apjs, 250, 2, \dodoi{10.3847/1538-4365/ab9d82}

\bibitem[{{Villaescusa-Navarro} {et~al.}(2021){Villaescusa-Navarro}, {Angl{\'e}s-Alc{\'a}zar}, {Genel}, {Spergel}, {Somerville}, {Dave}, {Pillepich}, {Hernquist}, {Nelson}, {Torrey}, {Narayanan}, {Li}, {Philcox}, {La Torre}, {Maria Delgado}, {Ho}, {Hassan}, {Burkhart}, {Wadekar}, {Battaglia}, {Contardo}, \& {Bryan}}]{navarro2021}
{Villaescusa-Navarro}, F., {Angl{\'e}s-Alc{\'a}zar}, D., {Genel}, S., {et~al.} 2021, \apj, 915, 71, \dodoi{10.3847/1538-4357/abf7ba}

\bibitem[{{Wang} {et~al.}(2016){Wang}, {Mo}, {Yang}, {Zhang}, {Shi}, {Jing}, {Liu}, {Li}, {Kang}, \& {Gao}}]{wang2016}
{Wang}, H., {Mo}, H.~J., {Yang}, X., {et~al.} 2016, \apj, 831, 164, \dodoi{10.3847/0004-637X/831/2/164}

\bibitem[{{White} \& {Rees}(1978)}]{white78}
{White}, S.~D.~M., \& {Rees}, M.~J. 1978, \mnras, 183, 341, \dodoi{10.1093/mnras/183.3.341}

\bibitem[{{Wu} \& {Kragh Jespersen}(2023)}]{WuJespersen23}
{Wu}, J.~F., \& {Kragh Jespersen}, C. 2023, arXiv e-prints, arXiv:2306.12327, \dodoi{10.48550/arXiv.2306.12327}

\end{thebibliography}

\end{document}